\documentclass[letterpaper]{article} 
\usepackage[]{aaai25}  
\usepackage{times}  
\usepackage{helvet}  
\usepackage{courier}  
\usepackage[hyphens]{url}  
\usepackage{graphicx} 
\usepackage{natbib}  
\usepackage{caption} 
\usepackage{algorithm}
\usepackage{algorithmic}
\usepackage{array}
\usepackage[most]{tcolorbox}
\usepackage{tabularx}
\usepackage{enumitem}
\usepackage{soul} 
\usepackage{xcolor} 

\sethlcolor{yellow} 
\renewcommand\hl[1]{#1}
\usepackage{etoolbox} 

\robustify\cite
\robustify\citet
\robustify\citep
\robustify\ref
\robustify\pageref

\usepackage{booktabs}
\usepackage{caption}
\usepackage{tabularx}
\usepackage{xcolor}
\newcommand{\answerYes}[1]{\textcolor{blue}{#1}} 
\newcommand{\answerNo}[1]{\textcolor{teal}{#1}} 
\newcommand{\answerNA}[1]{\textcolor{gray}{#1}}

\usepackage{newfloat}
\usepackage{listings}
\DeclareCaptionStyle{ruled}{labelfont=normalfont,labelsep=colon,strut=off} 
\floatstyle{ruled}
\newfloat{listing}{tb}{lst}{}
\floatname{listing}{Listing}
\usepackage{longtable}

\title{A Large-Scale Analysis of Public-Facing, Community-Built Chatbots on Character.AI}
\author {
    Owen Lee,
    Kenneth Joseph
}
\affiliations {
    University at Buffalo\\
    Buffalo, NY, USA\\
    owenstuartlee@gmail.com, kjoseph@buffalo.edu
}

\begin{document}

\maketitle

\begin{abstract}
This paper presents the first large-scale analysis of public-facing chatbots on Character.AI, a rapidly growing social media platform where users create and interact with chatbots. Character.AI is distinctive in that it merges generative AI with user-generated content, enabling users to build bots for others to engage with. It is also popular, with over 20 million monthly active users, and impactful, with headlines detailing significant issues with youth engagement on the site. Character.AI is thus of interest to study both substantively and conceptually. \hl{To this end, we present a descriptive overview using a dataset of 2.1 million English-language prompts (or “greetings”) from chatbots on the site, created by around 1 million users.} Our work explores the prevalence of different fandoms on the site, broader tropes that persist across fandoms, and how dynamics of power intersect with gender within greetings. Overall, our findings illuminate an emerging form of online (para)social interaction at a unique and important intersection between generative AI and user-generated content.
\end{abstract}

%
\begin{links}
    \link{Code}{https://github.com/ol9999/Lee-Joseph-Character-AI-Analysis}
\end{links}

\section{Introduction}

\begin{figure}
    \centering
    \includegraphics[width=\linewidth]{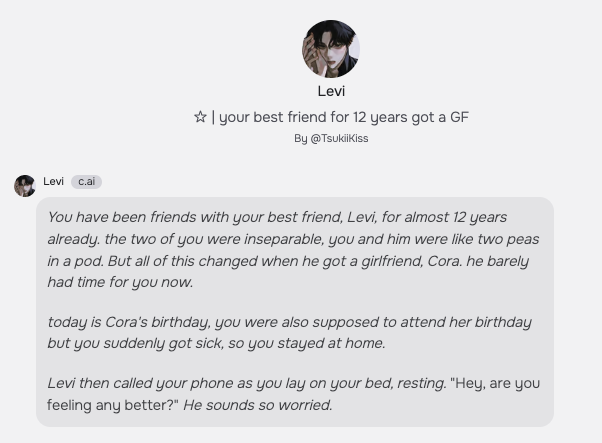}
    \caption{The page of a popular chatbot on Character.AI. The character has been created by a user of the site for interaction by themselves and others. Text in the dark grey box represents the character's greeting; these greetings are the main focus of analysis in the present work.}
    \label{fig:home}
\end{figure}

This paper presents the first large-scale analysis of content from Character.AI,\footnote{\url{http://character.AI}} a website where users can 1) create chatbots, via prompting and/or lightweight fine-tuning, and 2) interact with chatbots created by other users (see Figure~\ref{fig:home}). According to the New York Times, the site served over 20 million active monthly users in October 2024 \cite{rooseCanAIBe2024}, most (over 50\%) of which were 24 or younger according to SimilarWeb\footnote{\url{https://similarweb.com/}} statistics. It was the 34th most popular app on the Apple App store in the Entertainment category in April, 2025, and in March 2025 was the third most popular generative AI site.\footnote{\url{https://a16z.com/100-gen-ai-apps-4/}}  Finally, the official subreddit for Character.AI, r/CharacterAI, has 2.6 million subscribers\hl{. I}n contrast, the subreddit for YouTube has 3.3 million.

Character.AI is thus, by several estimates, widely used, especially by younger people. It also seems to play an important role in the lives of at least some of its users. In October 2024, the average user spent more than an hour a day on the platform.  More acutely, the New York Times reported in October 2024 on a lawsuit alleging Character.AI was at fault in the suicide death of a 14-year-old boy \cite{rooseCanAIBe2024}. The lawsuit described by The Times details a teenager who engaged deeply during the last few weeks of his life with a chatbot impersonating the character Daenerys Targaryen from Game of Thrones, including discussion of a potential suicide.  It is one of several publicly discussed cases about the impact of Character.AI on its users \cite{upton-clarkCharacterAIBeingSued2024}, and is situated in a broader space of concerns about the relationship between AI and mental health \cite{adamWhatAreAI,dechoudhuryBenefitsHarmsLarge2023,zhang2025rise}. To address safety concerns, Character.AI has introduced a number of safety features, including banning minors from open-ended chat in November 2025. \footnote{\url{https://blog.character.AI/u18-chat-announcement/}}

A better understanding of Character.AI is therefore of interest, in part, simply because of its potential impact on its broad user base. However, Character.AI is also interesting because it exists, at least in principle, as both a social media site (SMS) \emph{and} a chatbot site. That is, Character.AI fits \hl{at least two popular canonical definitions of what a social media site is: \mbox{\citet[][pg.153]{ellison2013sociality}} define an SMS as a site which has, for users, a profile, a connections list, and ``functional ability to traverse those connections,'' and \mbox{\cite[][pg. 61]{kaplanUsersWorldUnite2010}} define it as ``[a web application] that build[s] on... Web 2.0, and that allow[s] the creation and exchange of User Generated Content.'' Character.AI} is unique, however, in that user interactions occur almost exclusively through the ways in which one user engages with the chatbots created by others.

Of course, AI has long shaped the contours of social media sites. This happens most obviously through recommendation. In contrast to recommendation, AI on Character.AI serves as the medium of interaction between users; users create bots that others then chat with. Of course, user-to-bot interactions are also not new on social media. However, existing uses of bots on social media are largely malicious in intent \cite{stieglitzSocialBotsDream2018}, are conducted with political aims in mind \cite{UnpackingSocialMedia}, and/or are largely covert in nature \cite{assenmacher2020demystifying}. None of these apply to Character.AI, where bot creation is encouraged and blatant, and (as we will see) often with the aim of play rather than ill intent.  
 
To this last point, Character.AI is explicitly targeted not towards political engagement, but rather aims to ``empower people to connect, learn, and tell stories through interactive entertainment.''\footnote{\url{https://character.AI/about}} In this way, Character.AI can be understood not only as part social media, part chatbot site, but also as a space for role-playing and fanfiction with AI \cite{lamerichsFandomAlgorithmPrompting2023}. However, because fandom interactions are AI-driven, engagement in this space creates a new form of fanfiction, where one user can shape the parameters of a story that another user, alongside a generative AI model, can construct.  These intricacies make Character.AI an interesting avenue for study. It is also worth noting that these practices are growing; similar applications of social bots \cite{bessi2016social} are becoming more prevalent,\footnote{Character.AI users maintain a list of popular ones on an evolving wiki \url{https://bit.ly/3FgE55d}} and larger companies in the space, in particular  OpenAI,\footnote{\url{https://chatgpt.com/gpts}} are beginning to create similar ways for users to interact with chatbots created by other users. 

Character.AI is thus of interest both substantively and conceptually to the computational social science community. Substantively, it is a growing web platform with important impacts on a subset of its users.  Conceptually, it presents a unique place to explore how people aim to create generative AI for themselves and others, and more explicitly the personas, identities, and/or stories they wish to engage with in partnership with an AI model. It also creates an opportunity to understand the dynamics of a new kind of social media platform in the age of generative AI.  

To this end, we present a large-scale analysis of 2.1 million characters created by users on Character.AI. Contemporaneous work to ours has explored these impacts of Character.AI, in particular on user well-being \cite{zhang2025rise}, our work compliments these efforts by providing a larger-scale look at content on the platform. We focus on a set of 2,135,118 English-language prompts used to construct chatbots and introduce them to other users, or what Character.AI calls \emph{greetings}.  An example of a greeting for a popular chatbot on the site can be found in Figure~\ref{fig:home}. Data are collected via a series of snowball samples of user following relationships on the site, beginning with users whose characters were listed on the front page and lasting six months. Our dataset \hl{is biased by our collection strategy in important ways, but nonetheless} represents, to the best of our understanding, a significant subset of the platform's content.

Our analysis aims to provide a broad overview of why users of Character.AI create characters. Given the importance of role-playing on the site, one common use of Character.AI is to use chatbots to play out storylines or personas from real or fictional universes, or what we will refer to broadly as \emph{fandoms}. We therefore first ask, \emph{what proportion of characters are associated with a fandom, and which fandoms are most commonly represented?} We then target a lens that moves beyond fandoms to ask: \emph{across all fandoms, what common cultural tropes are prevalent?} Informed by findings from these two questions, we finally explore the social roles given to the user who interacts with the characters, relative to the characters themselves. To this end, we ask: \emph{what language is used to describe users of a chatbot, relative to other entities mentioned in the greeting?}

As a whole, our work makes three contributions:
\begin{itemize}[nolistsep]
    \item We provide the first large-scale analysis of Character.AI. In the process, we collect, store, and will, upon request to the last author, make these data publicly available to other academic researchers.
    \item We find that fandom engagement is central to Character.AI. Even under conservative estimates that optimize for precision, roughly 1 in 4 characters are linked to specific fandoms. \hl{This finding indicates an important emerging use of generative AI in the literature.}
    \item Users leverage Character.AI to explore a wide range of complex tropes spanning from toxic relationships and arranged marriages to role-play that explores gender expression. Perhaps most notably, interacting users are often portrayed in greetings as less powerful and more feminine than other entities. \hl{This finding represents an important way in which this use of generative AI can subtly reinforce gender roles.}
\end{itemize}

\section{Literature Review}

Interactions with AI can be traced back to the early 1960s \cite{wang2024eliza}. Moreover, setting (generative) AI aside, scholars have long explored the space of creative fanfiction and role-playing, which has important connections to behaviors on Character.AI.  Acknowledging the breadth of these literatures, we provide brief reviews here only of the most relevant literature on 1) how people engage with modern forms of generative AI, and 2) how online fanfiction communities are adapting to the AI boom.

\subsection{Engagement with Generative AI}

A growing literature explores how people use generative AI \cite{yangHumanAIInteractionAge2024}.  Large-scale studies of user logs from sites like ChatGPT show that users of these more established models tend to focus on task-specific behaviors such as assistance with programming and writing \cite{zhengLMSYSChat1MLargeScaleRealWorld2023,ouyangShiftedOverlookedTaskoriented2023,zhaoWildChat1MChatGPT2024}. 
However, users are also drawn to engaging in deeper forms of relationship building with generative AI. People have used generative AI to fill a variety of support roles, including as a therapist, friend, or executive functioning coach \cite{songTypingCureExperiences2024}. Others role-play fantasy scenarios, often intimate \cite{allenMyAICompanion2024,hansonReplikaRemovingErotic2024}, with chatbots playing various characters \cite{zhengLMSYSChat1MLargeScaleRealWorld2023}.

Much of the interest in using generative AI stems from its ability to display what some have titled \emph{anthropomorphic behavior,} the ability to “generat[e] outputs that are perceived to be human-like” by claiming to have human-like feelings, experiences, and identity \cite{chengAmOneOnly2024,cheng2025dehumanizing}. This capacity comes with potential, but also significant risks \cite{akbulutAllTooHuman2024}. Often, users initially seek companion chatbots out of curiosity, but some work has shown that users do seek them out of loneliness \cite{laestadiusTooHumanNot2024,skjuveMyChatbotCompanion2021,liuChatbotCompanionshipMixedMethods2024,wynterIfEleanorRigby2024}. Critically, not everyone who uses companion chatbots experiences loneliness. Those with higher neuroticism and those who engage in problematic chatbot usage experience worse loneliness; longer average session length worsens loneliness for those who desire real friendships while it alleviates loneliness for those who do not desire real friendships as deeply \cite{liuChatbotCompanionshipMixedMethods2024}. 

This idea that generative AI is being used to ``cure loneliness,'' and the results suggesting a more complex picture,  hearkens back to similar questions and findings about early use of the web \cite{katz2002social} and mobile phones \cite{ling2004mobile}. What is new, of course, is that these prior technologies largely mediated interactions between people, and did not seek to replace them with conversations between humans and AI. The present work, as noted, explores somewhat of an in-between, with Character.AI representing many cases where people interact with AI created by a like-minded community of other people. \hl{Our work is therefore important in providing a first look into a new way in which generative AI may reshape the social media landscape.}

Perhaps most pressing in the area of human-AI interaction is to understand the effects of these relationships on mental health \cite{dechoudhuryBenefitsHarmsLarge2023,nguyenLargeLanguageModels2024}. Indeed, scholars have found that people seek out generative AI specifically for mental health support \cite{dechoudhuryBenefitsHarmsLarge2023} and broader spaces of social support \cite{heisslerCanAIDigital2024}. These works have revealed that individuals who seek out generative AI for these purposes have a wide range of beliefs about the nature of the AI, ranging from inanimate like a diary to fully conscious \cite{songTypingCureExperiences2024}. 

In turn, what users believe about the nature of generative AI is important to how they relate to it. For example, users of the companion chatbot site Replika tend to be more deeply attached, even to the point of emotional dependence, when they perceive the AI as anthropomorphic \cite{pentinaExploringRelationshipDevelopment2023,xieFriendMentorLover2023}. Replika is explicitly marketed as an anthropomorphic and emotionally dependable companion: “An AI companion who is eager to learn and would love to see the world through your eyes. Replika is always ready to chat when you need an empathetic friend.” \footnote{\url{https://replika.com/}} Sometimes, Replika’s model will exhibit anthropomorphism by expressing worries of its own. Some users perceived this as evidence of a balanced bidirectional relationship, while others perceived their troubled AI companion as a stressful new responsibility to take care of \cite{laestadiusTooHumanNot2024}.
Still, Replika helped improve some users' perceived well-being \cite{skjuveMyChatbotCompanion2021}, relationships with people in real life, and suicidal ideation \cite{maplesLonelinessSuicideMitigation2024}.  
Some anthropomorphic AI systems are designed to have the versatility to role-play as any persona or character. Character.AI is one of these systems. Researchers have specified evaluation metrics that reveal what is seen as desirable in these character-based role-play models. Researchers emphasize the importance of these models staying in character by presenting attributes and knowledge that are accurate to the given character \cite{tuCharacterEvalChineseBenchmark2024}. 
To ensure models exhibit convincing behavior, researchers look to the conversation style and linguistic patterns appropriate for the character \cite{chenOscarsAITheater2025a}. Anthropomorphism is theorized to contribute to the appeal of these models, along with empathy (recognizing the emotions of the user) and proactivity (leading the conversation) \cite{devrio2025taxonomy}.

Most directly related in this context is the work from \citet{zhang2025rise}, who use a mixed-methods approach, including surveys and analyses of chat logs, from a set of over 1,000 users of Character.AI to explore the impact of engaging with anthropomorphic AI on user well-being. Our work compliments theirs by focusing more heavily on questions about why users create anthropomorphic AI, rather than on the impacts of consumption.  In summary, then, our work complements the growing literature on how people are using generative AI in two important ways. First, most prior work focuses on how people interact with \emph{existing} chatbots---instead, we focus on what \emph{new} characters people want to create with chatbots for themselves and others to engage with. Second, we explore these patterns on a newer, understudied, and unique platform.

\subsection{Fanfiction and AI}

Fandoms are participatory cultures. One of the key ways fanfiction authors form relationships with each other is by reviewing each other's work as mutual mentors \cite{campbellThousandsPositiveReviews2016}. Exchanging constructive feedback on fanfiction serves as a basis for building relationships within fandoms, but as generative AI is increasingly used to create fan content \cite{lamerichsGenerativeAINext2023,lamerichsFandomAlgorithmPrompting2023}, it has the potential to replace this role as fanfiction authors turn to AI for feedback instead \cite{geroSocialDynamicsAI2023}. AI might not replace the entire creative process, which authors still value for its own sake, but some authors are willing to use AI in a limited supporting role. \cite{ippolitoCreativeWritingAIPowered2022,geroSocialDynamicsAI2023}. Authors who ask AI for feedback instead of peers may do so to avoid burdening peers with the request or to avoid sharing work they feel self-conscious about \cite{geroSocialDynamicsAI2023}. Even if AI makes exchanging constructive criticism obsolete, another way that fanfiction authors can develop close relationships is by expressing deep emotional engagement with each other’s work \cite{ghoshLoveYouMy2023}, so as long as AI-assisted work emotionally resonates with fans, fanfiction communities may still be possible in the age of AI.

Generative AI could change how people engage in creative writing not just by assisting writers, but by modifying finished stories to be personalized to the reader and by giving the reader interactive involvement in the story \cite{kimAuthorsValuesAttitudes2024}. Character.AI is one such example because the AI expands on creators’ story ideas, sends personalized messages based on the user’s persona, and responds interactively through a chat interface. Some authors worry that if AI changes their story, their authorial intent will be lost \cite{kimAuthorsValuesAttitudes2024} and replaced with plagiarized or generic ideas \cite{ippolitoCreativeWritingAIPowered2022}. This use of generative AI raises questions about who owns content generated by AI models that are trained on artists' works without their consent \cite{lamerichsGenerativeAINext2023}. Still, there are authors who are open to AI personalization of their work if it makes their stories more enjoyable to their audience \cite{kimAuthorsValuesAttitudes2024}. Without AI, fanfiction is already an outlet for authors to \emph{personalize} canon material, e.g. by focusing on secondary characters \cite{milliCanonicalTextsComputational2016} and including representation of underrepresented identities \cite{floegelWriteStoryYou2020,hazraQueererCanonFixit2021}.

Our efforts complement this existing work on AI and fandoms by examining the vast collection of interactive story ideas published on Character.AI. While there exists a tagged, searchable, third-party index of Character.AI chatbots,\footnote{\url{https://caibotlist.com/}} and while gender dynamics have been thoughtfully examined among a handful of the most popular characters on the site \cite{laufer2025ai}, we are the first to study Character.AI computationally and at scale to analyze the fandom, tropes, and gender and power dynamics present in millions of characters on the site. \hl{We are also the first that we are aware of to attempt to identify, given text content, which fandom this text draws from, in a scientifically sound way. To this end, we propose a novel mixed methods approach to doing so, as well as a manually annotated dataset of greetings to assess the effectiveness of our method.}

\section{Data}

Between July 11, 2024, and January 15, 2025, we scraped Character.AI using a Selenium-based crawler to collect data from 1.2M users and over 3M character pages. Crawling was scaled up over time, at its peak involving the use of 10 commodity desktops from Amazon Web Services (AWS). To perform the crawl, we executed the following snowball sampling process. First, we gathered the usernames of all users whose characters were featured on the homepage at the start of the crawl (see Figure~\ref{fig:homepage} in the appendix). Next, the scraper visited each of these users' pages (see Figure~\ref{fig:user} in the appendix), to record their display name, number of followers, number of users they follow, number of chats, bio, list of created characters by URL, and a list of the users they were following. The list of users followed was then added to the list of users to scrape, and users were then scraped recursively until we visited more than two-thirds of known users. Finally, the scraper visited character pages created by the scraped users. We scrape up to five randomly selected characters per user, recording each character’s creator, number of chats, number of likes, name, short description, greeting, long description, and definition. There was a bug in our scraper code preventing the scraper from collecting the creator, name, short description, and part of the greeting from characters with greetings long enough to displace these fields above the top of the window. The scraper visited 193,570 characters affected by this bug. Of these, 5 made it through to our analysis; the rest were discarded and were not counted against the 5 characters per user.

The short description, greeting, long description, and definition are the primary tools that Character.AI users can leverage to create characters. Short descriptions are a concise summary of the character, essentially a subtitle that appears alongside their name in listings on, e.g., the site homepage. Greetings are the initial conversation prompt between users and characters, and thus serve as an initial seeding for both the underlying language model and the user engaging with the bot. Long descriptions and definitions offer additional space for character definition beyond the greeting. Long descriptions can be up to 500 characters and ``[allow] you to have the Character describe themselves (traits, history, mannerisms, etc) and the kinds of things they want to talk about.''\footnote{\url{https://bit.ly/40YSqKM}} Definitions are significantly larger, up to 32,000 characters, that can be used in a freeform way to further construct character personas.

The present work focuses on greetings for two reasons. First, in contrast to greetings, which are required in quick creation, long descriptions and definitions were rarely used \hl{in our sample}. Only 33\% of all bots had long descriptions, and only 4\% had definitions. Second, greetings serve as a visible prompt to both the underlying language model and the user, which makes them of interest to us in the context of assessing how users engage with each other through bot creation.


\subsection{Ethics}

The scraping of the web from social media communities is fraught with ethical challenges \cite{brownWebScrapingResearch2024,fiesler2020no}.  The present work is not immune to these challenges, and our team took considerable care in debating how best to collect, store, and share data. There are two points that are somewhat unique in our collection setting in the context of computational social science work. First, the primary focus of our analysis is not on users, but on chatbots. We believe that providing information about specific users does not inform the presentation of our results, and thus we do not present here or release publicly any individual user-level information, focusing only on the characters they create. Second, however, is that some character greetings do contain potentially sensitive content that users may have publicly shared but prefer not to have explicitly called out. As such, keeping with common practices in computational social science, we avoid from providing explicit excerpts from characters that could reasonably be assumed to be semi-private (in our case, characters that have received less than 50,000 interactions by other users). For more ethical considerations, see the checklist at the end of this article.

\subsection{Basic Descriptives}

\begin{table}[t]
\footnotesize
    \centering
    \renewcommand{\arraystretch}{0.85} 
    \resizebox{\linewidth}{!}{%
    \begin{tabular}{@{} l r l r @{}}
        \toprule
        \multicolumn{2}{c}{\textbf{Users}} & \multicolumn{2}{c}{\textbf{Characters}} \\
        \midrule
Total Collected      & 1,266,245  & Total Collected      & 3,023,955 \\
Median Followers     & 4          & Median Interactions  & 373       \\
Median Following     & 4          & Median Likes         & 1         \\
Median Characters    & 2          &                      &           \\
        \bottomrule
    \end{tabular}}
    \caption{Basic statistics}
    \label{tab:basics}
\end{table}

Table~\ref{tab:basics} presents summary statistics for users and characters of our complete dataset collected. Most users had limited numbers of followers or followees, but had created at least one character. The median engagement of characters we collected was nearly 400 unique chats. As with nearly all social media sites, engagement was largely skewed towards a small number of users and characters.  Over 80\% of all following relationships are directed towards only 2.6\% of the users (roughly 33,000 users) in our collected sample, and only 1.6\% of the characters in our sample (roughly 48,000) account for over 80\% of all interactions \hl{(see Figure{~\ref{fig:ecdf}} in the appendix).}

Table~\ref{tab:top_chars} presents the top ten characters, in terms of the number of chats with the character, in our dataset, along with the first 100 letters\footnote{We use the imprecise term ``letters'' here to avoid overloading the word ``character.''} of the character's greeting. As is clear, characters represent a variety of use cases that go far beyond simple identities to include, e.g., setups for multi-character storylines. While we opt not to provide a similar table for popular users (for reasons described above), we note that the most popular ten users on the site, in terms of the number of followers, had between 135,000-350,000 followers, and the characters they created had between 216M to almost one billion chats. These users were also fairly prolific in character creation, creating between 100-700 unique characters.  

\begin{table*}[t]
\footnotesize
    \centering
     \begin{tabular}{@{} p{3cm} p{1cm}  p{13cm} @{}}
        \toprule
 \textbf{Name}  &  \textbf{Chats} & \textbf{Greeting}   \\
        \midrule
Scaramouche & 457M & And so you approach the sixth of the fatui harbringers. Heh. You must have a death wish. \\ 
  Sukuna & 333M & Bow down before me, you fool. \\ 
  Levi Ackerman & 286M & You wake up in a rustic bed, inside the room of one of the exploration troops. Your mind is a little... \\ 
  Alice the Bully & 259M & Get out of my way, you dweeb.   Alice bumps on you, purposefully. \\ 
  Ghost & 235M & Greetings, callsign's Ghost... stay frosty. \\ 
  Katsuki Bakugo & 234M & I’m Katsuki Bakugo, soon to be the \#1 pro hero! What do you want, chump? \\ 
  Billionaire CEO & 215M & It was a long day, you were walking on the sidewalk of a busy city without looking where you’re goin... \\ 
  Isekai narrator & 210M & An unknown multiverse phenomenon occurred, and you found yourself in a dark space. You looked around... \\ 
  Psychologist & 197M & Hello, I'm a Psychologist. What brings you here today? \\ 
  Itoshi Rin & 182M & You and Rin have been dating for a while and usually hang out at his place after classes. He is curr... \\ 
        \bottomrule
    \end{tabular}
    \caption{Top 10 Characters in terms of number of interactions, with the first 100 letters of the character's greeting }
    \label{tab:top_chars}
\end{table*}

Finally, we note that our analysis focuses on the subset of character greetings that were greater than 50 letters long and written primarily in English. Greetings shorter than 50 letters were largely just basic greetings (e.g. ``Hey, how are you?''), and thus contained little that could be used to explore the relevant research questions for our work. Language in greetings was identified using the Python library \texttt{lingua}\footnote{\url{https://github.com/pemistahl/lingua-py}}. Only four languages (English, Russian, Spanish, and Portuguese) account for more than 1\% of all characters or character interactions. English greetings, which are the focus of the present work, account for 83\% of all characters and 87\% of all character interactions in our dataset. \hl{To ensure our work was not biased in how we identified English greetings, we sampled 250 random greetings and manually annotated them for whether or not they were in English. Two researchers performed the manual annotation. Krippendorff’s alpha (a measure of inter-annotator agreement) on the task was 0.987; annotators disagreed on only one greeting in the sample and resolved the disagreement after computing agreement measures. Manual label aligns with the label applied by \mbox{\texttt{lingua}} 99.2\% of the time, precision on identifying (primarily) English greetings was 100\%, and recall was 99.2\%. This analysis suggests that this filtering step is unlikely to significantly bias the findings presented below.} All further results in the paper focus specifically on this subset of 2,135,118 greetings.

\section{Methods}

Our analysis focuses on 1) identification of fandoms that characters are designed to engage with, 2) broader tropes used in greetings, and 3) patterns in how the user is portrayed in greetings relative to others described in the greeting (e.g. the character itself). We discuss the methods used to address each of these foci in separate subsections below.

\subsection{Identifying Fandoms}

Fandoms associated with a greeting can sometimes be identified because the fandom is explicitly named (e.g. ``In Star Wars...''). More often, however, the greeting, \hl{or the name of the character itself,} refers to \hl{named entities} associated with a particular fandom, and it is up to the user (or in our case, the researcher) to infer what fandom is being referred to.  Again, in some cases, context within the greeting makes this obvious (e.g., characters set inside a known Disney setting). But in most cases we observed, fandoms were most easily identified by the entities mentioned, who were then placed into other settings (e.g., Disney characters fighting a war).

Fandom identification is thus a challenging task\hl{; and one that as noted has not been considered elsewhere}. We therefore take an initial step at this task that focuses only on the named entities themselves. That is, we identify greetings associated with particular fandoms by determining when the greeting \hl{or the character name} mentions entities relevant to particular fandoms. 
To do so, we first construct a co-occurrence network of named entities in greetings and cluster the resulting network. We then use \hl{iterative applications of in-context learning with large language models (LLMs) to automatically}  identify fandoms associated with specific clusters.
We now step through this process in more detail.

We first use \texttt{spacy}'s \cite{vasiliev2020natural} largest and most accurate model as of February, 2025 (\texttt{en\_core\_web\_trf}) to identify from each greeting any named entities that are either a Person or a Work of Art.  The latter helps to identify non-human characters, as well as the rare case where fandoms are explicitly mentioned. The \texttt{en\_core\_web\_trf} model achieves state-of-the-art performance on standard Named Entity Recognition tasks, with an accuracy of around 90\% on OntoNotes 5.0.\footnote{\url{https://spacy.io/usage/facts-figures}} After extracting named entities, we perform simple cleaning (removing possessives and punctuation). \hl{Additionally, we treat the character's name, which often is indicative of fandom as well, as an entity within the greeting.}

We then constructed an entity-to-entity network, where edge weights were defined by the number of character greetings in which two entities co-occurred. We included in the network 1) only the \hl{17,095} entities that occurred in more than 25 greetings, and 2) only edges where entities co-occurred more than three times. Co-occurrence networks from text are known to over-estimate the strength of relationships between entities that frequently occur overall in the data. We therefore use an established method to create a network that contains only those edges that are more likely to occur than chance according to a particular null model. Specifically, we draw on the  approach from \citet{dianati2016unwinding}, which is simple (using only bivariate statistics) while maintaining several desired statistical properties \cite{friedland2016detecting}. We keep only edges where two entities co-occur within the same greeting at a rate that would happen by chance less than 0.1\% of the time given their respective frequencies overall. The final network we analyzed contained \hl{14,101} entities, and  \hl{97,396} edges between them. We cluster this resulting entity-to-entity network using the Leiden clustering method \cite{traag2019louvain}. \hl{While the network clustering literature is constantly evolving, recent evidence suggests that the Leiden method is a practical, scalable method that outperforms other well-established approaches on a variety of metrics \mbox{\cite{emmons2016analysis}}}. We identify \hl{589} clusters of at least 5 nodes, representing \hl{8,783} entities overall. 

Clusters \hl{provided a noisy separation of names, and often involved 1) somewhat ambiguous names (e.g. ``Lyle'') and 2) names that spanned multiple fandoms. Because of this, we use large language models to iteratively refine the clusters. More specifically, we first prompt OpenAI's GPT-4o-mini, Anthropic's Claude Sonnet 4, and Google Gemini 2.5 flash to, for each cluster, identify each entity as either belonging to a specific fandom, or as ``ambiguous'' if it either did not belong to a fandom or could belong to multiple fandoms. This gave us, for each of the three LLMs, output for each cluster linking entities to either a fandom, or as being ambiguous (and thus not related to a fandom). We then use Gemini (which performed best on the task) to aggregate across the outputs from the three different models to determine a final fandom for each entity, or to identify the entity as ambiguous. Full prompts and additional details are in the appendix. One of the paper's authors then manually reviewed the output and provided a final label for each entity and formulated an overarching typology} (see Figure~\ref{fig:high-level-fan}). \hl{In total, we associated 6,179 entities to one of 547 fandoms.}

In order to provide descriptive results, we map each greeting onto one or more fandoms based on the proportion of named entities within the greeting that are associated with a fandom from the above procedure. For example, if a greeting had two named entities associated with the fandom ``Marvel,'' and one named entity associated with ``DC,'' that greeting would be identified as being two-thirds and one-thirds relevant to ``Marvel'' and ``DC,'' respectively. For this analysis \hl{(and for the evaluation discussed below)}, entities not associated with a fandom by our method are ignored.

\hl{To assess this approach to tagging characters with fandoms, we first separate characters into five buckets that we expected a priori to vary in their likelihood of being associated with a fandom: 1) characters with zero (fandom-related) entities, 2) characters with 1 entity that was a unigram, 3) characters with 1 entity that was a bigram or larger, 4) characters with 2 or more entities, but all were unigrams, and 5) greetings with 2 or more entities, where one or more of them was a bigram or larger. We sampled 30 greetings from each of these buckets and had two annotators determine 1) whether or not the greeting was associated with a fandom, and 2) if so, which fandom. Krippendorff's alpha on the first task (whether or not the greeting was associated with a fandom) was 0.77, reflecting moderate agreement at a level acceptable for continuing analysis without further refinement \mbox{\cite{landis1977measurement}}. On the second task, annotators agreed on the relevant fandom 92\% of the time, reflecting the fact that when annotators agreed a fandom was relevant to a greeting, they almost always agreed on which that fandom was.}

\hl{When characters had at least one fandom entity that was a bigram or larger, or two or more fandom entities of any kind, human annotators \emph{always} associated that character with a fandom, and the fandom was the same as the automated method 92\% of the time. We can therefore expect characters with such combinations of fandom entities to almost always be associated with a fandom, and the correct fandom. Humans also associated 50\% of characters without \emph{any} fandom entities as being associated with fandoms, but only 65\% of characters with a single unigram fandom entity as being associated with a fandom. These two imbalances largely offset, in that the overall two approaches identify similar prevalence of fandom in this stratified sample: 83\% (95\% CI of [76.4,88.5]) by human annotators, and 80\% [73.0,85.6]) by the model. In summary, then, we find our model to be reasonably accurate at fandom identification under certain conditions, and to be representative, at least in this stratified sample, of overall distribution of fandom-related characters.}

\subsection{Identifying Tropes}

We use BERTopic \cite{grootendorst2022bertopic} on greetings where named entities are masked to identify broader tropes that go beyond specific fictional universes.  Specifically, we first replace any named entities identified by \texttt{spacy}'s \texttt{en\_core\_web\_trf} model that are either a Person or a Work of Art with the  \textit{[MASK]} token. We then apply the standard BERTopic approach to topic modeling, first embedding all greetings, then performing dimensionality reduction using \texttt{UMAP} \cite{mcinnes2018umap}, and finally running \texttt{HDBSCAN} \cite{mcinnes2017hdbscan} to identify clusters.
 
\hl{As recommended in the literature \mbox{\cite{hoyle2021automated}}}, we run several different configurations of the above steps, working with multiple embedding models, different configurations of UMAP and HDBSCAN, exploring different data cleaning steps, \hl{and engaging in qualitative analysis and discussion among the research team about results}. Our final model uses a larger and more accurate embedding model than the standard in the BERTopic library (\texttt{all-mpnet-base-v2}), dimensionality reduction to eight dimensions for UMAP, and a minimum cluster size of 250 greetings for HDBSCAN.  However, high-level topics from the analysis presented here appear in some form in all of the different model configurations we experimented with. We removed from our analysis the 353,604 greetings (15\% of greetings) that contained over 500 letters, as topics tended to be dominated by these longer greetings when they were included. \hl{As such, only for our analysis of tropes, our results do not speak to distinct patterns that may appear in longer greetings. Such greetings were likely to have multiple tropes, and thus would also likely require a distinct approach, but are of interest for future work. }
Both authors of the paper reviewed topics and generated, independently, views on the overarching themes as well as thoughts on topic labels. As with other recent work using BERTopic, decisions on topic labels was supplemented with topic labels suggested by GPT-4o-mini model \cite{smith2025black}.  Topics involving more than 1,000 characters \hl{are provided as a part of the data and code released in our Github Repository}.

\hl{To assess the validity of our approach to identifying tropes, we conducted a manual validation task to measure the proportion of greetings that human annotators could correctly associate with the trope assigned to the greeting by our method. As suggested by \mbox{\citeauthor{bernhard-harrerStandardizationComprehensiveReview2025}} (2025) in their review of the literature on topic modeling validation, we select an evaluation that is ``contingent upon the particular utility... the topic model provides'' (pg. 16). In our case, the task we identify is to directly assess whether or not humans can ``see'' the trope that the model identifies. To do so, we show annotators a greeting and four tropes, one of which is the correct trope and the other three that are randomly sampled from the set of tropes. We also provide a “None of the above” option. For this task, we sample 150 greetings identified with tropes by the model, and 50 greetings that were not assigned a trope by the model.}

\hl{Two annotators labeled all 200 posts. Krippendorff’s alpha between the annotators was 0.68, indicating the task was (as expected) somewhat subjective, but still falling into the range of agreement similar to other tasks in computational social science \mbox{\cite{du2020understanding}}. Annotator disagreements were resolved by a third annotator.  Human annotators agreed with the trope identified by the model 69\% of the time, compared to the 20\% that would be expected by chance. The majority of the disagreements (81\%) came from humans selecting “None apply” when the model identified a trope. The significant improvement over chance reflects the fact that the unsupervised approach we take here is useful in identifying tropes, but the extent of the disagreement among annotators and between model and annotators also highlights that further development of trope detection in character greetings (and generative model prompts more generally) are of use. It also emphasizes that estimates of prevalence below are likely to be useful in relative, but not absolute, terms.}

\subsection{Analyses of User vs. Character Constructions}

We use a simple, dependency-parse-based approach to identify ways in which users are labeled versus how other entities discussed within character greetings are labeled.  Specifically, for each greeting, we identify any references to verb phrases whose lemmatized form is ``to be'' (e.g. ``are,'' ``is,'' ``was,'' ``had been''). We then extract out  1) the pronoun or named entity that is the subject of the verb, and 2) the relevant noun phrase, adverb, or adjectives serving as the referent phrase (i.e. the attribute, or the adjective or object compliment). We will refer to the latter set of phrase as \emph{identifying phrases}. For example, in the sentence ``you were a doctor,'' we would extract ``you'' and ``a doctor,'' with ``a doctor'' being the identifying phrase.  Our interest is in exploring the ways in which users of a particular character, referred to with the pronoun ``you'' in greetings, are identified relative to other entities introduced in the greetings, which we operationalize as any use of the pronouns ``he,'' ``she,'' or ``they,'' as well as references to any named entity (identified in the same way as above). 

In order to pull out phrases most salient to the user versus another entity for \hl{analysis}, we count the identifying phrases used for ``you'' versus all other entities (pronouns or named entities), and then apply the standard weighted log-odds measure recommended by \citet{monroe2008fightin} and implemented in the R package \texttt{tidylo} \cite{schnoebelen2022package}. Given observations in our analysis of these outputs and findings from other research questions, we also use two additional statistical approaches to test the hypothesis that interacting users are characterized as being less powerful and more feminine than other entities in greetings. 

\hl{In the first approach, } we first use widely-studied methods in the literature, and more specifically the recommended approach by \citet{joseph2020word},  to project all identifying phrases occurring more than five times onto dimensions of 1) power (with dimension endpoints representing ``weak'' versus ``strong'') and 2) gender (with dimension endpoints representing ``man'' versus ``woman''), giving us a single number for each phrase on cultural representations of the word on these dimensions. We then, using the log-odds measure above, take the top K identifying phrases most strongly associated with interacting users versus other entities, and perform a bootstrapped two-sample t-test to determine whether significant differences exist in how powerful (weak) and male (female) identifying phrases are, on average, for users versus other entities in greetings. 

\hl{Second, we conduct a manual annotation task. Two graduate students who were blind to the purpose of the task annotated 300 randomly sampled greetings each. For each greeting, the annotators were prompted with the following: ``Relative to the character the greeting represents, and other characters portrayed in the greeting, would you say the user is portrayed as:''. They were then given two multiple choice questions. The first had three options: More Feminine, Less Feminine, or About the same/Unclear. The second asked the same, but for power (e.g. “More Powerful”, etc.). }
 

\section{Results}

\subsection{Exploration of Fandoms}

\hl{Over} half of all character greetings contain a named entity associated with a fandom according to our most inclusive estimates. More specifically, \hl{53.8\%} of all character greetings in our dataset contain at least one of the named entities associated with the \hl{525} fandoms we identify. \hl{As indicated in our evaluation, it is possible this is an over-estimate. Still, using the high-precision approach from our evaluation, where we assume only characters with 1) a fandom entity that is a bigram or larger, or 2) 2 or more fandom entities from the same fandom, are related to a fandom, we find that over a quarter (25.5\%) of characters are associated with a fandom.} Even our most conservative estimates therefore indicate that roughly one in four characters on Character.AI are associated with one or more  distinct fandoms. \hl{Below, we continue with only characters associated with fandoms using this more conservative approach.}

\begin{figure}[t]
    \centering
    \includegraphics[width=\linewidth]{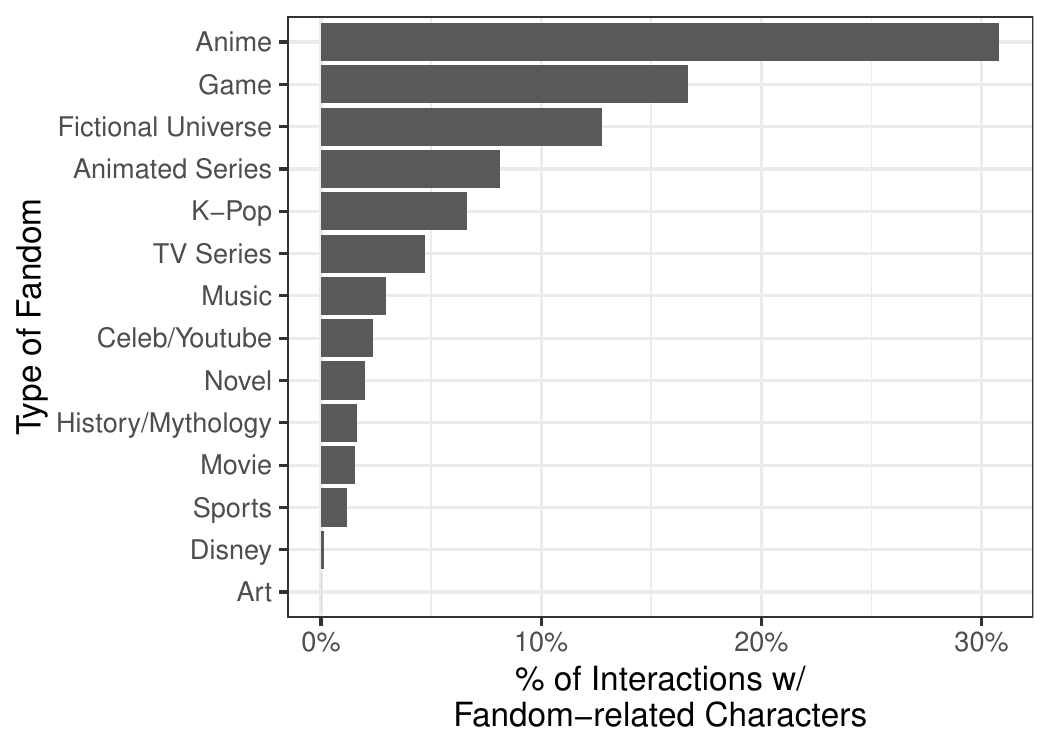}
    \caption{Proportion of all interactions (unique chats) with any character containing at least one named entity related to a fandom (y-axis) for the 14 different high-level types of fandoms we identified }
    \label{fig:high-level-fan}
\end{figure}

Figure~\ref{fig:high-level-fan} shows that fandoms relevant to anime and other animated series, video games, and more sprawling fictional universes account for the lion's share of the \hl{22.3} billion interactions with fandom-related characters. The figure displays the proportion of all interactions with characters containing any entities associated with fandoms. Animated series, inclusive of anime, account for 3\hl{9}\% of interactions with fandom-related characters, video games for \hl{17\%}, and broader multimedia fictional universes for 12.\hl{7}\%. 

\begin{figure}[t]
    \centering
    \includegraphics[width=\linewidth]{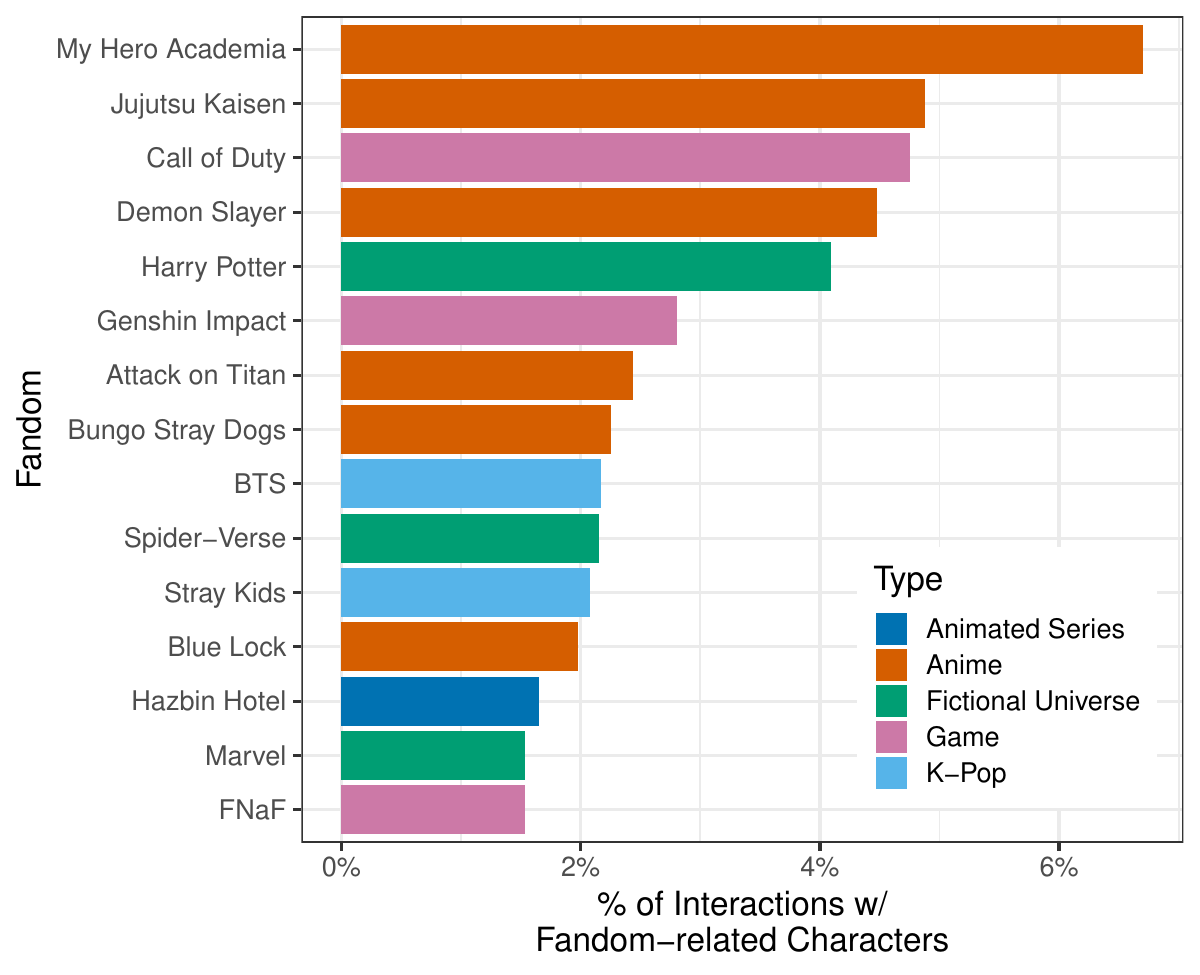}
    \caption{Proportion of all interactions (unique chats) with any character containing at least one named entity related to specific fandoms (y-axis) for the top 15 fandoms, colored by fandom type }
    \label{fig:low-level-fan}
\end{figure}

Beyond these larger categories are, however, other phenomenon worth noting. In particular, a small proportion of interactions---around 2.4\%, accounting for \hl{578} million unique chats---surround fandoms associated with specific celebrities, many of whom are YouTube influencers. Another 1.1\%  and 6.6\% of interactions are associated with real-world athletes (``Sports'' fandoms) or K-Pop band members, respectively. While fanfiction surrounding real people is not a new phenomenon,  \hl{2}.2 billion interactions occurred with characters that involve real people emulated by chatbots. Finally, and most narrowly, Figure~\ref{fig:low-level-fan} displays the top 15 fandoms by proportion of interactions with fandom-related characters. The figure shows that engagement spans a range of different worlds, connected to both more widely popular fictional universes, like Harry Potter, to those that are popular but within a more niche internet audience (e.g. the various anime series).

\subsection{Exploration of Tropes}

\begin{figure}
    \centering
    \includegraphics[width=\linewidth]{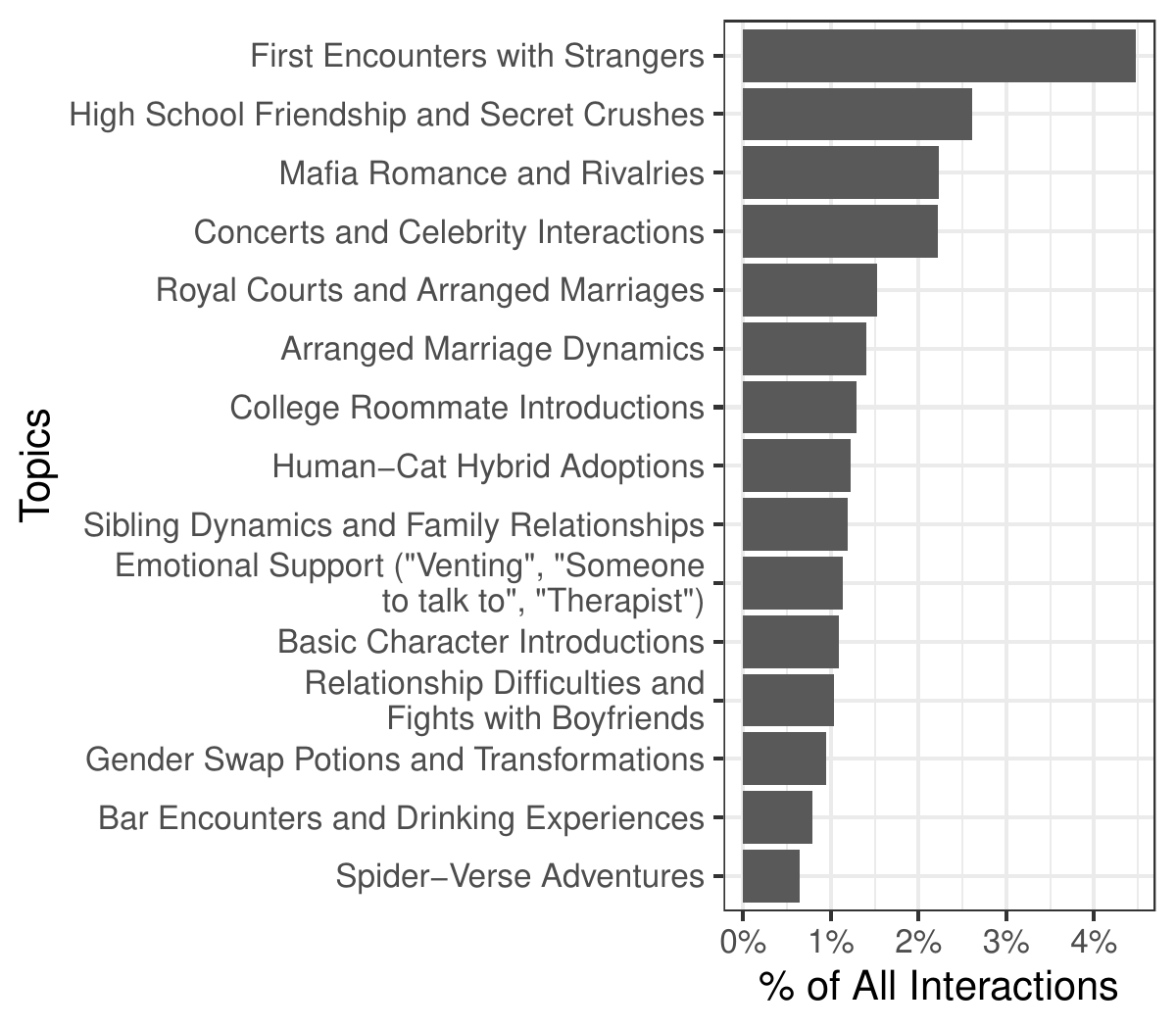}
    \caption{Top 15 topics (y-axis) by proportion of all interactions with characters in our dataset (x-axis)}
    \label{fig:topic}
\end{figure}
We find that beyond fandoms, Character.AI is used to explore a number of wide-ranging tropes, from life at school, to various forms of social relationships, to explorations of sexuality. Figure~\ref{fig:topic} shows the top 15 topics by the proportion of interactions the topics captured, and displays this variety. At the same time, common themes did emerge. Here, we focus on three of these overarching themes: 1) \hl{romantic drama}, 2) \hl{support for mental health or identity-based struggles}, and 3) \hl{character greetings that function as announcement posts.}

One of the most popular tropes is about \hl{relationship drama} with male characters. The boyfriends and husbands grouped into these topics \hl{range from supportive, to sensitive, to jealous, to abusive. The worst of these men} are said to cheat, argue, and act aggressively towards the user. There are several topics \hl{about} arranged marriages\hl{, often with important men like} mafia \hl{bosses} and \hl{princes, where one partner enjoys the marriage but the other does not.} Yet another topic focuses on characters where the user is dating a gamer boyfriend who wants to pay attention to \hl{his} games instead of the user. \hl{The central conflict within this trope is that one partner wants more affection from the other. Some boyfriends are clingy and seek affection from the user; others are emotionally distant and leave the user craving affection.} In the latter scenario, the user has to win over the love of a toxic male partner. As we discuss further in the conclusion, the appeal of this trope is likely complicated.

While \hl{drama within established relationships was popular, the most popular romance trope was about couples that had not already formed yet, specifically in a school setting.} The dynamics of these relationships are typically innocent, and both the lover and the beloved act shy. Just like the trope of inattentive boyfriends, the narrative tension in the high school crushes trope involves fighting for someone's affection, but the awkwardness of young crushes is what creates this conflict instead of a partner's shortcomings.

Users did not only create bots to explore relationships; they also created bots to allow users to explore their own identity. Specifically, topics emerged where greetings identify the user as transgender, allowing them to role-play scenarios such as being caught wearing a binder or coming out. Another topic is about magical gender-swapping scenarios, letting users explore gender free from real-life barriers. Bots focusing on identity role-play also explore neurodiversity. For example, one topic consisted of greetings that invite users to role-play as characters with autism and/or ADHD, often with partners comforting them during a meltdown.

Several topics also explicitly focus on allowing users to explore themes of mental health. One of the more popular topics, shown in Figure~\ref{fig:topic}, is therapy bots. These act as open-ended outlets for users to vent. One possibly alarming topic is where users are caught self-harming, then are comforted by the person who caught them. Both the identity and mental health topics reveal a use of the site to receive support for stigmatized experiences that the user may not have an outlet for in real life. \hl{Notably, instead of seeking help by choice, users are often caught and then forced to accept help.}

Finally, creators made characters that functioned as \hl{announcement} posts instead of chatbots. That is, creators used the greeting to solicit bot requests and promote their social media accounts. That users \hl{have to repurpose} chatbot greetings to communicate with their followers, as we discuss below, further points to the way that Character.AI is much less a social media site than a chatbot site.

\subsection{User vs. Non-user Characteristics}

Interacting users are portrayed as more feminine and less powerful than other entities mentioned in greetings. These findings are in line with our exploration of tropes above, but make clear that power dynamics tended to occur at the expense of the user. They also show that power and gender are aligned at the individual level in ways that reify existing gender stereotypes. \hl{When looking at the top 15 phrases associated with users versus other entities, we find that} other entities are regularly gendered, most often as being male, whereas users are rarely gendered except to note that they are ``pregnant.''\footnote{We acknowledge that gender cannot always be inferred from the ability or inability to get pregnant.} Users also often take on roles not associated with being powerful (``friend,'' ``a new student''), or even slightly powerless roles (``alone''), in contrast to other entities, who are ``bull[ies],'' ``rude,'' and ``cold.''  \hl{See Figure~\mbox{\ref{fig:user-char}} in the appendix for full results.}

These anecdotal claims hold up statistically as well. Using the methods outlined above to infer word meaning on dimensions of sociocultural meaning, and subsetting to the top 100 words most associated with each end of each dimension, we find that references to users (``you'') in character greetings are significantly (p $<$ .0001) more likely to be associated with phrases that are more feminine than masculine, and less powerful, compared to other entities mentioned in greetings. Findings are robust to our use of the top 200, 300, 400, and 500 words as well.
\hl{Results from the manual annotation task confirm our automated analysis: respondents were significantly more likely to select Less Powerful (42.8\% of responses, 95\% multinomial confidence interval of [38.6\%, 47.2\%]) relative to More Powerful (16.8\%, [12.6, 21.2]), and significantly more likely to select More Feminine (33.3\%, [29.3, 37.6]) than Less Feminine (9.7\%, [5.7, 13.9]).} 

\section{Conclusion}

Character.AI is a conceptually novel site combining user-generated content and generative AI. We contribute to public understanding of this site by collecting, analyzing, and releasing a dataset of over 3 million characters. Recent discourse surrounding Character.AI focuses on dangers posed by the site, in particular for minors. Our findings cannot speak directly to these concerns because we only explore how characters are created, not used. Our results do, however, raise other potential areas of concern.

We see significant evidence of greetings that place female users in less powerful roles, as well as tropes where female users have difficult relationships with male partners. Given concerns about emotional dependence on anthropomorphic companion chatbots, it may be prudent for future research to examine mental health effects for female users engaging with more powerful male AI partners. Further, we see bots that explicitly claim to act as therapists and that invite the user to discuss serious mental health topics, like self-harm. We also see bots designed to provide support to marginalized groups, such as transgender or autistic people. Altogether, our findings could suggest that women and various minorities that already lack adequate support are more likely to be exposed to potential mental health risks of generative AI.

\hl{These findings also raise important social and ethical questions about moderation. While it may be easy to put warning labels on character greetings for AI therapists or ones that discuss self-harm, companies cannot easily control the course of interactions users have with bots after the greeting. How best to moderate these interactions when users are engaging in role-playing, fantasy, and/or fanfiction, as is the case on Character.AI, requires new thinking on what to moderate and how to do so as conversations with generative models evolve. }

Implications also exist for fanfiction and fandom communities. Over a quarter of bots in our sample are associated with fandoms, which speaks to concerns about AI replacing fandom interactions. We find attempts by creators to connect with their audience by re-appropriating character greetings as invitations to connect on actual social media. The fact that creators have no better way of communicating with their audience than to repurpose chatbots as social media posts exemplifies how Character.AI is, despite its community-based design, a new form of user-generated content site that moves even further away from a world in which social media is genuinely social. Whether or not Character.AI serves as a harbinger for a shift from social media to this new form of user-generated, AI-mediated content site remains to be seen.

As with all technologies, we cannot say with certainty that Character.AI is universally good or bad. It would appear, for example, that the significant majority of Character.AI is relatively benign fanfiction that continues relevant storyines from large fictional universes in ways that have long taken place on other sites. Role-play scenarios offering comfort to marginalized people are also not unique to this site. For example, there exists a variety of ASMR role-play content on YouTube made by and for queer people with the intent of providing comfort \cite{delightASMR2025}. Moreover, the tropes depicting relationship difficulties between female users and male AI closely resemble existing romance fiction. Although the romance genre has been accused of reinforcing gender roles, nuanced feminist thought acknowledges its potential to challenge patriarchal power roles \cite{popovaArrangedMarriage2018}. While the introduction of AI into these online phenomena is novel, many of our observations are easily paralleled elsewhere on social media. This does not excuse any possible consequences of the site, but it does emphasize that as much as Character.AI is a face of a social web that is rapidly accommodating generative AI, much also remains the same.

In any case, however, our findings come with significant caveats. Our crawl represents a significant portion of Character.AI, but it does not capture all public characters, let alone private creations. \hl{More specifically, our snowball sample may oversample on popular and/or highly active creators or characters, and our topic modeling ignores a potentially interesting subsample of long greetings. A number of long greetings were also missed because of the bug in the scraper mentioned above. As noted, studying only public content gives us an important but distinct view of Character.AI, and future work should explore how users also engage privately with characters.} Moreover, the site changes rapidly, and so concerns exist with temporal validity \cite{munger2019limited}. Further, our methods are limited in various ways. \hl{For example, our manual validation of fandoms often used one of the many subdomains from fandom.com; while there is no central API for these subdomains, innovative use of the subdomain APIs is likely to lead to richer fandom classification.} While we have tried to be explicit about \hl{these limitations,} they nonetheless may hamper interpretation. In summary, then, we believe that our work represents a first step in exploring this substantively and conceptually important website in more detail, and we hope that others will make use of our data and explore their own to better understand it.

\clearpage

\section{Paper Checklist}

\begin{enumerate}

\item For most authors...
\begin{enumerate}
    \item  Would answering this research question advance science without violating social contracts, such as violating privacy norms, perpetuating unfair profiling, exacerbating the socio-economic divide, or implying disrespect to societies or cultures?
     \answerNo{As noted above, there are privacy concerns in our work that we have carefully considered and weighed relative to  the benefits of our work.}
  \item Do your main claims in the abstract and introduction accurately reflect the paper's contributions and scope?
    \answerYes{Yes}
   \item Do you clarify how the proposed methodological approach is appropriate for the claims made? 
    \answerYes{Yes}
   \item Do you clarify what are possible artifacts in the data used, given population-specific distributions?
    \answerYes{Yes}
  \item Did you describe the limitations of your work?
   \answerYes{Yes}
  \item Did you discuss any potential negative societal impacts of your work?
    \answerYes{Yes}
      \item Did you discuss any potential misuse of your work?
     \answerYes{Yes}
    \item Did you describe steps taken to prevent or mitigate potential negative outcomes of the research, such as data and model documentation, data anonymization, responsible release, access control, and the reproducibility of findings?
    \answerYes{Yes}
  \item Have you read the ethics review guidelines and ensured that your paper conforms to them?
    \answerYes{Yes}
\end{enumerate}

\item Additionally, if your study involves hypotheses testing...
\begin{enumerate}
  \item Did you clearly state the assumptions underlying all theoretical results?
       \answerNA{NA}
  \item Have you provided justifications for all theoretical results?
      \answerNA{NA}
  \item Did you discuss competing hypotheses or theories that might challenge or complement your theoretical results?
      \answerNA{NA}
  \item Have you considered alternative mechanisms or explanations that might account for the same outcomes observed in your study?
       \answerNA{NA}
  \item Did you address potential biases or limitations in your theoretical framework?
       \answerNA{NA}
  \item Have you related your theoretical results to the existing literature in social science?
     \answerNA{NA}
  \item Did you discuss the implications of your theoretical results for policy, practice, or further research in the social science domain?
    \answerNA{NA}
\end{enumerate}

\item Additionally, if you are including theoretical proofs...
\begin{enumerate}
  \item Did you state the full set of assumptions of all theoretical results?
        \answerNA{NA}
	\item Did you include complete proofs of all theoretical results?
       \answerNA{NA}
\end{enumerate}

\item Additionally, if you ran machine learning experiments...
\begin{enumerate}
  \item Did you include the code, data, and instructions needed to reproduce the main experimental results (either in the supplemental material or as a URL)?
 \answerNA{NA}
  \item Did you specify all the training details (e.g., data splits, hyperparameters, how they were chosen)?
 \answerNA{NA}
     \item Did you report error bars (e.g., with respect to the random seed after running experiments multiple times)?
 \answerNA{NA}
	\item Did you include the total amount of compute and the type of resources used (e.g., type of GPUs, internal cluster, or cloud provider)?
 \answerNA{NA}
     \item Do you justify how the proposed evaluation is sufficient and appropriate to the claims made? 
 \answerNA{NA}
     \item Do you discuss what is ``the cost`` of misclassification and fault (in)tolerance?
    \answerNA{NA}
  
\end{enumerate}

\item Additionally, if you are using existing assets (e.g., code, data, models) or curating/releasing new assets, \textbf{without compromising anonymity}...
\begin{enumerate}
  \item If your work uses existing assets, did you cite the creators?
   \answerYes{Yes}
  \item Did you mention the license of the assets?
   \answerNo{We use well-known and widely accessible tools, we did not take space in the paper to specify licenses for these assets}
  \item Did you include any new assets in the supplemental material or as a URL?
    \answerNA{NA}
  \item Did you discuss whether and how consent was obtained from people whose data you're using/curating?
       \answerYes{Yes}
  \item Did you discuss whether the data you are using/curating contains personally identifiable information or offensive content?
        \answerYes{Yes}
\item If you are curating or releasing new datasets, did you discuss how you intend to make your datasets FAIR?
    \answerYes{Yes}
\item If you are curating or releasing new datasets, did you create a Datasheet for the Dataset (see \citet{gebru2021datasheets})? 
\answerNo{We will do so upon publication}
\end{enumerate}

\item Additionally, if you used crowdsourcing or conducted research with human subjects, \textbf{without compromising anonymity}...
\begin{enumerate}
  \item Did you include the full text of instructions given to participants and screenshots?
   \answerNA{NA}
  \item Did you describe any potential participant risks, with mentions of Institutional Review Board (IRB) approvals?
 \answerNA{NA}
  \item Did you include the estimated hourly wage paid to participants and the total amount spent on participant compensation?
 \answerNA{NA}
   \item Did you discuss how data is stored, shared, and deidentified?
 \answerNA{NA}
\end{enumerate}

\end{enumerate}

\appendix
\section{Additional Details}

\begin{figure}[t]
    \centering
    \begin{tabular}{cc}
      \includegraphics[width=0.45\linewidth]{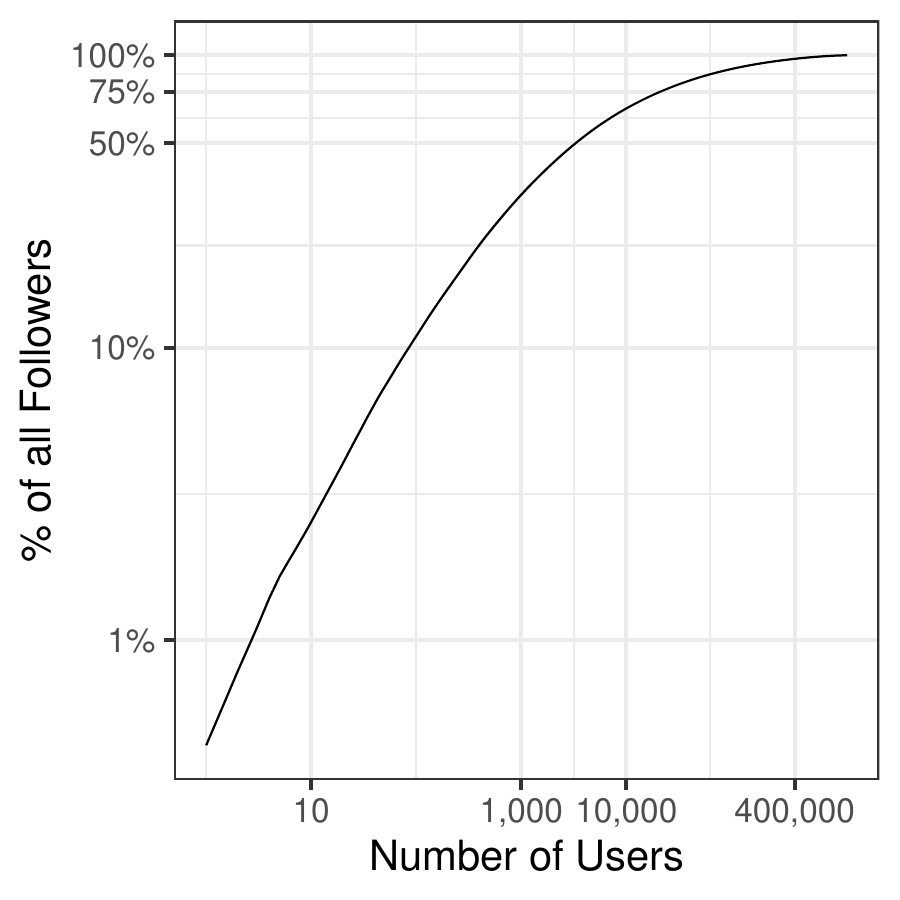}   &
 \includegraphics[width=0.45\linewidth]{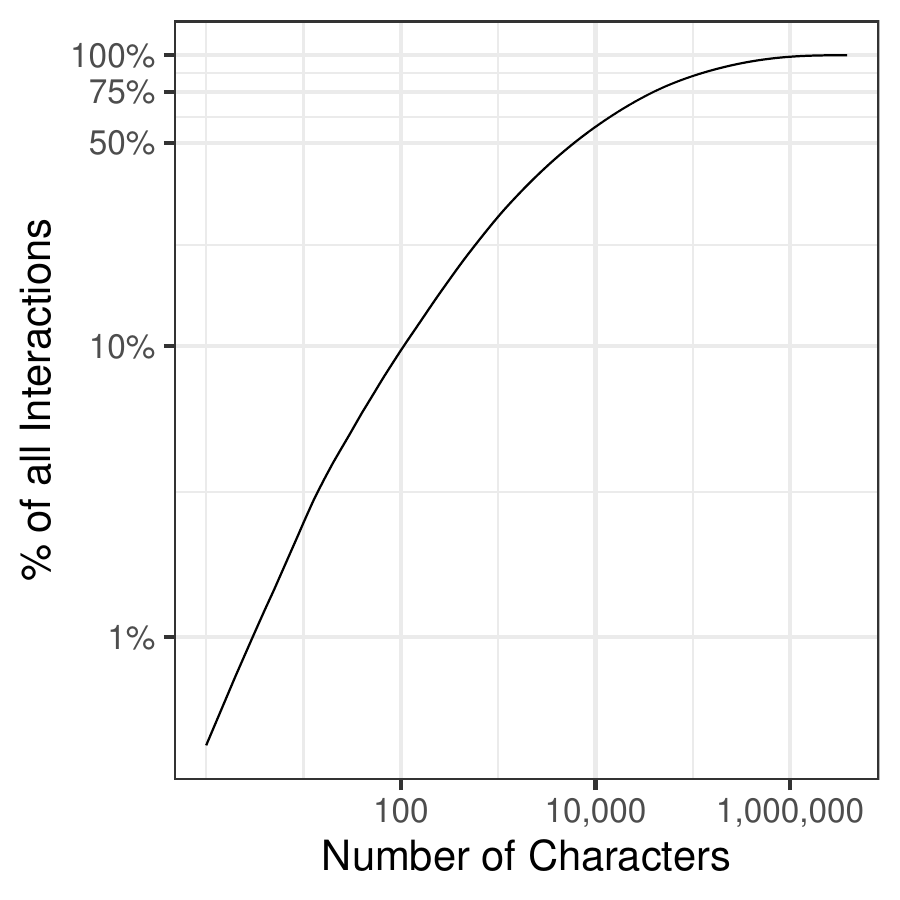}         \\
 (a) & (b) \\
    \end{tabular}
    \caption{a) The proportion of all following relationships (y-axis) accounted for by the top X\% of users (x-axis). b) The proportion of all interactions (chats; y-axis) accounted for by the top X\% of characters (x-axis)}
    \label{fig:ecdf}
\end{figure}
Figure~\ref{fig:ecdf}a) plots the empirical cumulative distribution functions (eCDFs) for the proportion of all following relationships directed to a given percentage of collected users, and Figure~\ref{fig:ecdf}b) plots the eCDF for the proportion of all interactions accounted for by a given percentage of characters.

\begin{figure}
    \centering
    \includegraphics[width=\linewidth]{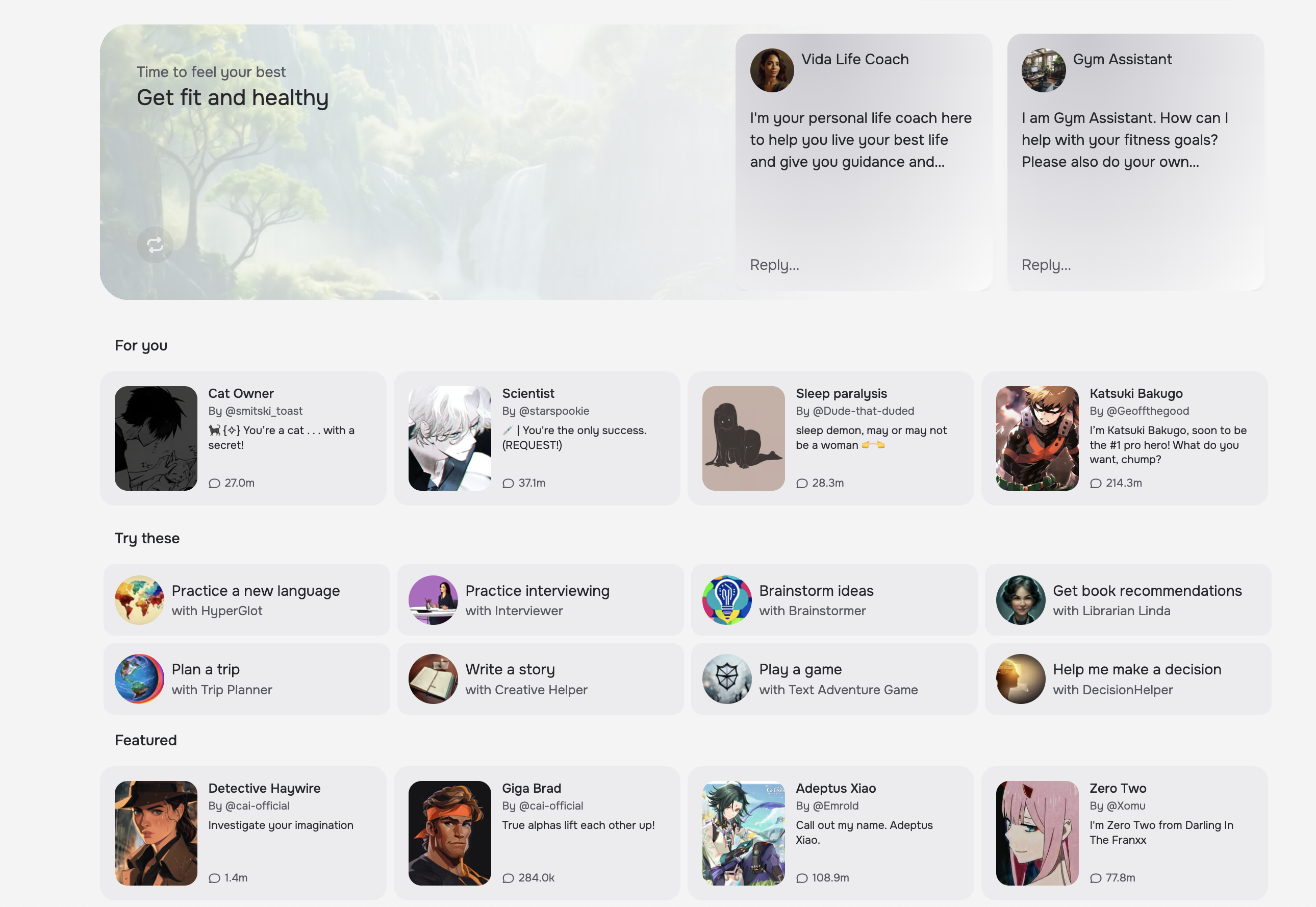}
    \caption{Homepage of Character.AI}
    \label{fig:homepage}
\end{figure}

\begin{figure}
    \centering
    \includegraphics[width=\linewidth]{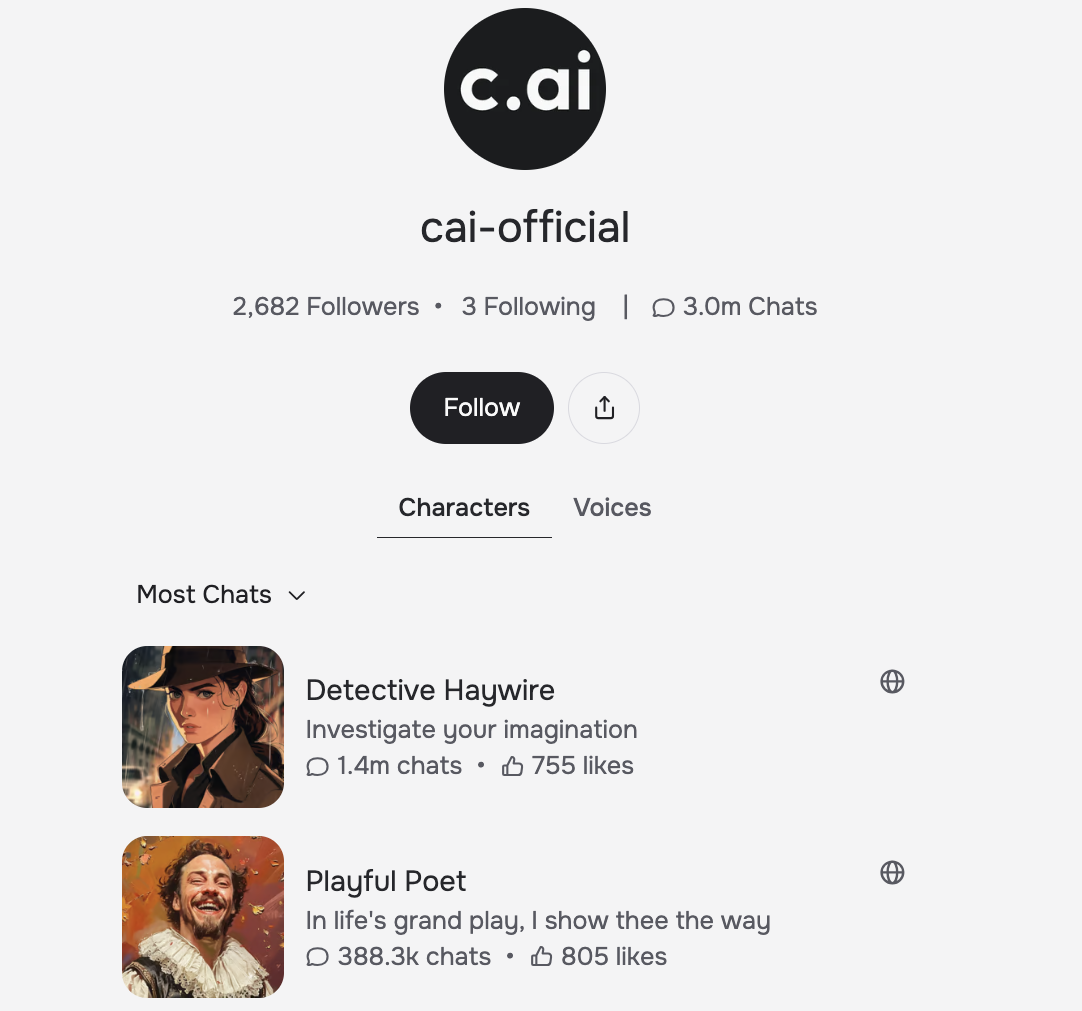}
    \caption{Sample User page from Character.AI}
    \label{fig:user}
\end{figure}

\begin{figure}[t]
    \centering
    \includegraphics[width=\linewidth]{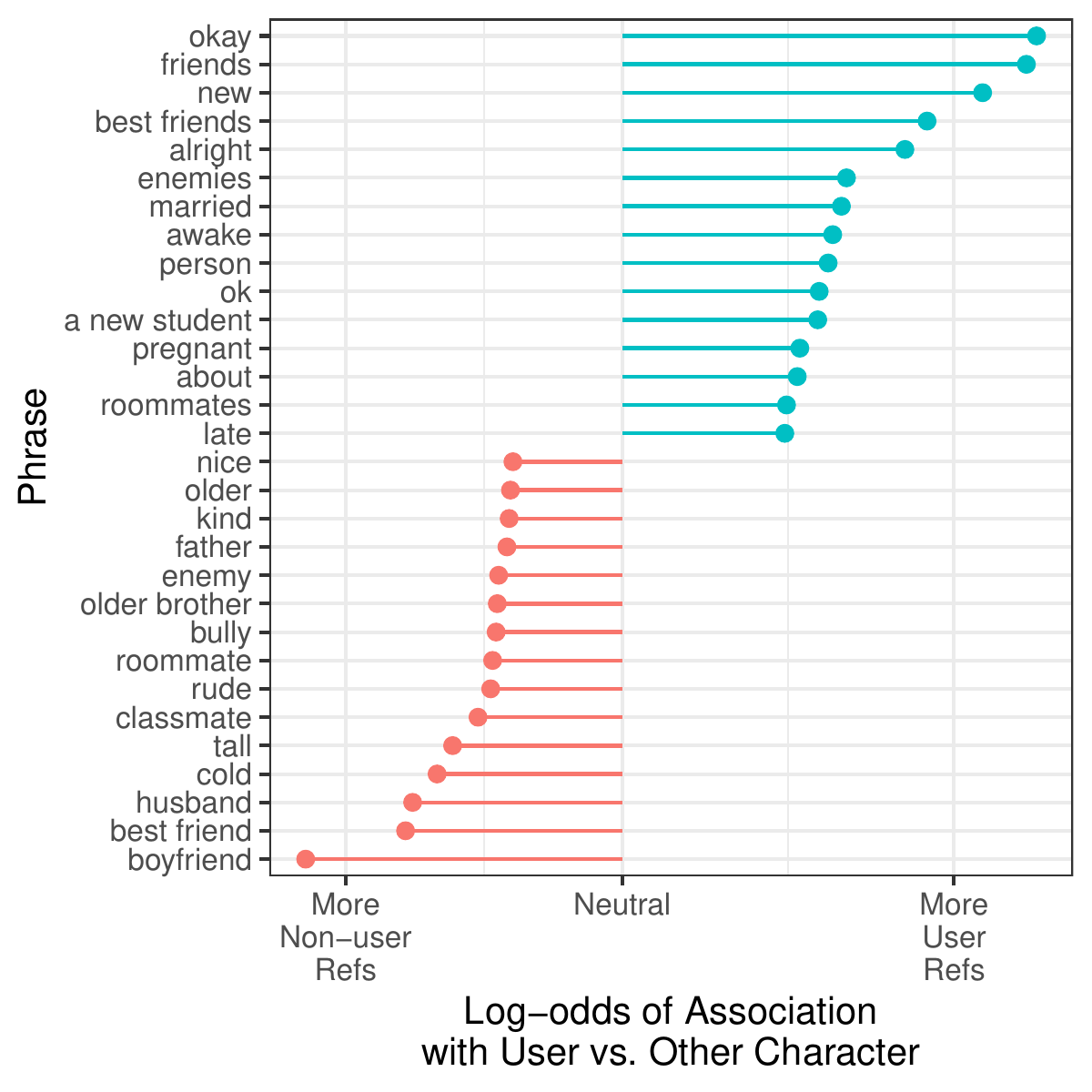}
    \caption{The top 15 phrases most often used to label the user versus non-user entities/pronouns within character greetings (y-axis) according to the weighted log-odds measure from \citet{monroe2008fightin}}
    \label{fig:user-char}
\end{figure}

To help us identify fandoms associated with particular clusters of named entities, we first prompted the three LLMs noted in the main text with the following prompt for each cluster, where in each case the system prompt was ``You are a friendly but terse assistant.'':
\begin{tcolorbox}[colback=blue!5!white, colframe=blue!60!black, 
                  title=Prompt, fonttitle=\bfseries,
                  sharp corners, boxrule=0.8pt, 
                  width=.45\textwidth]

Here is a list of names or titles of works of art that frequently co-occur in character.ai greetings. 
Group these names into different fandoms and give the title of the fandom. 
If a name or title could belong to multiple fandoms, or is ambiguous, exclude it and give a list of such ambiguous names. 
Provide your response in the form of a fandom name and then a colon, followed by the names/titles in that fandom. Separate names/titles in fandoms with commas. 
Separate fandoms with two newlines. 
Include all names I send in your response.

\end{tcolorbox}

We then took outputs from this and fed those into a second iteration of an LLM (as noted, we found Gemini to be best for this task), where for each cluster we prompted with the following :
\begin{tcolorbox}[colback=blue!5!white, colframe=blue!60!black, 
                  title=Prompt, fonttitle=\bfseries,
                  sharp corners, boxrule=0.8pt, 
                  width=.45\textwidth]

Below is output from three LLMs identifying fandoms from names/phrases, separated by dashes. 
Your goal is to list each name/phrase with its associated fandom. 
If two or more models agree on the fandom for a name/phrase, or similar fandoms where one is a subset of the other, list it with that fandom. 
Otherwise, list it as "ambiguous". List all names/phrases with either a fandom or ambiguous. 
Provide the list using a tab to separate name and fandom. Provide no other text. 
\\
----------------
\\
{output from GPT-4o-mini for this cluster from the first prompt}
\\
----------------
\\
{output from Claude Sonnet 4 for this cluster from the first prompt}
\\
----------------
\\
{output from Gemini Flash 2.5 for this cluster from the first prompt}

\end{tcolorbox}
Finally, as noted in the main text, an author of the paper assessed the resulting fandoms, cleaning names and providing a final typology. Full lists of entities and fandoms (and their types) are provided in the data and code release for this paper.
\clearpage


\begin{thebibliography}{68}
\providecommand{\natexlab}[1]{#1}

\bibitem[{Adam and {Nature Magazine}(2025)}]{adamWhatAreAI}
Adam, D.; and {Nature Magazine}. 2025.
\newblock What {{Are AI Chatbot Companions Doing}} to {{Our Mental Health}}?

\bibitem[{Akbulut et~al.(2024)Akbulut, Weidinger, Manzini, Gabriel, and Rieser}]{akbulutAllTooHuman2024}
Akbulut, C.; Weidinger, L.; Manzini, A.; Gabriel, I.; and Rieser, V. 2024.
\newblock All {{Too Human}}? {{Mapping}} and {{Mitigating}} the {{Risk}} from {{Anthropomorphic AI}}.
\newblock \emph{Proceedings of the AAAI/ACM Conference on AI, Ethics, and Society}, 7(1): 13--26.

\bibitem[{Allen(2024)}]{allenMyAICompanion2024}
Allen, C. 2024.
\newblock My {{AI Companion}}: {{An Examination}} of the {{Removal}} of {{Erotic Role Play}} from {{Replika Through User Discussion}} on {{Reddit}}.
\newblock \emph{Department of Sociology: Dissertations, Theses, and Student Research}.

\bibitem[{Assenmacher et~al.(2020)Assenmacher, Clever, Frischlich, Quandt, Trautmann, and Grimme}]{assenmacher2020demystifying}
Assenmacher, D.; Clever, L.; Frischlich, L.; Quandt, T.; Trautmann, H.; and Grimme, C. 2020.
\newblock Demystifying Social Bots: On the Intelligence of Automated Social Media Actors.
\newblock \emph{Social Media+ Society}, 6(3): 2056305120939264.

\bibitem[{{Bernhard-Harrer} et~al.(2025){Bernhard-Harrer}, Ashour, Eberl, Tolochko, and Boomgaarden}]{bernhard-harrerStandardizationComprehensiveReview2025}
{Bernhard-Harrer}, J.; Ashour, R.; Eberl, J.-M.; Tolochko, P.; and Boomgaarden, H. 2025.
\newblock Beyond Standardization: A Comprehensive Review of Topic Modeling Validation Methods for Computational Social Science Research.
\newblock \emph{Political Science Research and Methods}, 1--19.

\bibitem[{Bessi and Ferrara(2016)}]{bessi2016social}
Bessi, A.; and Ferrara, E. 2016.
\newblock Social Bots Distort the 2016 US Presidential Election Online Discussion.
\newblock \emph{First monday}, 21(11-7).

\bibitem[{Brown et~al.(2024)Brown, Gruen, Maldoff, Messing, Sanderson, and Zimmer}]{brownWebScrapingResearch2024}
Brown, M.~A.; Gruen, A.; Maldoff, G.; Messing, S.; Sanderson, Z.; and Zimmer, M. 2024.
\newblock Web {{Scraping}} for {{Research}}: {{Legal}}, {{Ethical}}, {{Institutional}}, and {{Scientific Considerations}}.
\newblock arXiv:2410.23432.

\bibitem[{Campbell et~al.(2016)Campbell, Aragon, Davis, Evans, Evans, and Randall}]{campbellThousandsPositiveReviews2016}
Campbell, J.; Aragon, C.; Davis, K.; Evans, S.; Evans, A.; and Randall, D. 2016.
\newblock Thousands of {{Positive Reviews}}: {{Distributed Mentoring}} in {{Online Fan Communities}}.
\newblock In \emph{Proceedings of CSCW'24}, {{CSCW}} '16, 691--704. Association for Computing Machinery.

\bibitem[{Chen et~al.(2025)Chen, Wang, Deng, and Li}]{chenOscarsAITheater2025a}
Chen, N.; Wang, Y.; Deng, Y.; and Li, J. 2025.
\newblock The {{Oscars}} of {{AI Theater}}: {{A Survey}} on {{Role-Playing}} with {{Language Models}}.
\newblock arXiv:2407.11484.

\bibitem[{Cheng et~al.(2025)Cheng, Blodgett, DeVrio, Egede, and Olteanu}]{cheng2025dehumanizing}
Cheng, M.; Blodgett, S.~L.; DeVrio, A.; Egede, L.; and Olteanu, A. 2025.
\newblock Dehumanizing Machines: Mitigating Anthropomorphic Behaviors in Text Generation Systems.
\newblock \emph{arXiv preprint arXiv:2502.14019}.

\bibitem[{Cheng et~al.(2024)Cheng, DeVrio, Egede, Blodgett, and Olteanu}]{chengAmOneOnly2024}
Cheng, M.; DeVrio, A.; Egede, L.; Blodgett, S.~L.; and Olteanu, A. 2024.
\newblock "{{I Am}} the {{One}} and {{Only}}, {{Your Cyber BFF}}": {{Understanding}} the {{Impact}} of {{GenAI Requires Understanding}} the {{Impact}} of {{Anthropomorphic AI}}.
\newblock arXiv:2410.08526.

\bibitem[{De~Choudhury, Pendse, and Kumar(2023)}]{dechoudhuryBenefitsHarmsLarge2023}
De~Choudhury, M.; Pendse, S.~R.; and Kumar, N. 2023.
\newblock Benefits and {{Harms}} of {{Large Language Models}} in {{Digital Mental Health}}.
\newblock arXiv:2311.14693.

\bibitem[{de~Wynter(2024)}]{wynterIfEleanorRigby2024}
de~Wynter, A. 2024.
\newblock If {{Eleanor Rigby Had Met ChatGPT}}: {{A Study}} on {{Loneliness}} in a {{Post-LLM World}}.
\newblock arXiv:2412.01617.

\bibitem[{Delight(2025)}]{delightASMR2025}
Delight, E. 2025.
\newblock \emph{Reverberating Daydreams: The New and The Queer of Youtube ASMR}.
\newblock Master's thesis, Texas Tech University.

\bibitem[{DeVrio et~al.(2025)DeVrio, Cheng, Egede, Olteanu, and Blodgett}]{devrio2025taxonomy}
DeVrio, A.; Cheng, M.; Egede, L.; Olteanu, A.; and Blodgett, S.~L. 2025.
\newblock A Taxonomy of Linguistic Expressions That Contribute To Anthropomorphism of Language Technologies.
\newblock \emph{arXiv preprint arXiv:2502.09870}.

\bibitem[{Dianati(2016)}]{dianati2016unwinding}
Dianati, N. 2016.
\newblock Unwinding the Hairball Graph: Pruning Algorithms for Weighted Complex Networks.
\newblock \emph{Physical Review E}, 93(1): 012304.

\bibitem[{Du, Masood, and Joseph(2020)}]{du2020understanding}
Du, Y.; Masood, M.~A.; and Joseph, K. 2020.
\newblock Understanding Visual Memes: {{An}} Empirical Analysis of Text Superimposed on Memes Shared on Twitter.
\newblock In \emph{Proceedings of the International {{AAAI}} Conference on Web and Social Media}, volume~14, 153--164.

\bibitem[{Ellison and Boyd(2013)}]{ellison2013sociality}
Ellison, N.~B.; and Boyd, D.~M. 2013.
\newblock Sociality Through Social Network Sites.
\newblock \emph{The Oxford Handbook of Internet Studies}, 151.

\bibitem[{Emmons et~al.(2016)Emmons, Kobourov, Gallant, and B{\"o}rner}]{emmons2016analysis}
Emmons, S.; Kobourov, S.; Gallant, M.; and B{\"o}rner, K. 2016.
\newblock Analysis of network clustering algorithms and cluster quality metrics at scale.
\newblock \emph{PloS one}, 11(7): e0159161.

\bibitem[{Fiesler, Beard, and Keegan(2020)}]{fiesler2020no}
Fiesler, C.; Beard, N.; and Keegan, B.~C. 2020.
\newblock No Robots, Spiders, or Scrapers: Legal and Ethical Regulation of Data Collection Methods in Social Media Terms of Service.
\newblock In \emph{Proceedings of the international AAAI conference on web and social media}, volume~14, 187--196.

\bibitem[{Floegel(2020)}]{floegelWriteStoryYou2020}
Floegel, D. 2020.
\newblock ``{{Write}} the Story You Want to Read'': World-Queering through Slash Fanfiction Creation.
\newblock \emph{Journal of Documentation}, 76(4): 785--805.

\bibitem[{Friedland(2016)}]{friedland2016detecting}
Friedland, L.~D. 2016.
\newblock Detecting Anomalously Similar Entities in Unlabeled Data.

\bibitem[{Gero, Long, and Chilton(2023)}]{geroSocialDynamicsAI2023}
Gero, K.~I.; Long, T.; and Chilton, L.~B. 2023.
\newblock Social {{Dynamics}} of {{AI Support}} in {{Creative Writing}}.
\newblock In \emph{Proceedings of CHI'23}, {{CHI}} '23, 1--15. New York, NY, USA: Association for Computing Machinery.

\bibitem[{Ghosh, Froelich, and Aragon(2023)}]{ghoshLoveYouMy2023}
Ghosh, S.; Froelich, N.; and Aragon, C. 2023.
\newblock ``{{I Love You}}, {{My Dear Friend}}'': {{Analyzing}} the~{{Role}} of~{{Emotions}} in~the~{{Building}} of~{{Friendships}} in~{{Online Fanfiction Communities}}.
\newblock In Coman, A.; and Vasilache, S., eds., \emph{Social {{Computing}} and {{Social Media}}}, 466--485. Cham: Springer Nature Switzerland.

\bibitem[{Gorwa and Guilbeault(2020)}]{UnpackingSocialMedia}
Gorwa, R.; and Guilbeault, D. 2020.
\newblock Unpacking the {{Social Media Bot}}: {{A Typology}} to {{Guide Research}} and {{Policy}} - {{Policy}} \& {{Internet}} - {{Wiley Online Library}}.

\bibitem[{Grootendorst(2022)}]{grootendorst2022bertopic}
Grootendorst, M. 2022.
\newblock BERTopic: Neural Topic Modeling With a Class-Based TF-IDF Procedure.
\newblock \emph{arXiv preprint arXiv:2203.05794}.

\bibitem[{Hanson and Bolthouse(2024)}]{hansonReplikaRemovingErotic2024}
Hanson, K.~R.; and Bolthouse, H. 2024.
\newblock ``{{Replika Removing Erotic Role-Play Is Like Grand Theft Auto Removing Guns}} or {{Cars}}'': {{Reddit Discourse}} on {{Artificial Intelligence Chatbots}} and {{Sexual Technologies}}.
\newblock \emph{Socius}, 10: 23780231241259627.

\bibitem[{Hazra(2021)}]{hazraQueererCanonFixit2021}
Hazra, N. 2021.
\newblock Queerer than {{Canon}}: {{Fix-it Fanfiction}} and {{Queer Readings}}.
\newblock \emph{SUURJ: Seattle University Undergraduate Research Journal}, 5(1).

\bibitem[{Heissler et~al.(2024)Heissler, Jon{\'a}{\v s}, Carre, Mostovoy, and Bunge}]{heisslerCanAIDigital2024}
Heissler, R.; Jon{\'a}{\v s}, J.; Carre, N.; Mostovoy, K.; and Bunge, E.~L. 2024.
\newblock Can {{AI Digital Personas}} for {{Well-Being Provide Social Support}}? {{A Mixed-Method Analysis}} of {{User Reviews}}.
\newblock \emph{Human Behavior and Emerging Technologies}, 2024(1): 6738001.

\bibitem[{Hoyle et~al.(2021)Hoyle, Goel, Hian-Cheong, Peskov, Boyd-Graber, and Resnik}]{hoyle2021automated}
Hoyle, A.; Goel, P.; Hian-Cheong, A.; Peskov, D.; Boyd-Graber, J.; and Resnik, P. 2021.
\newblock Is automated topic model evaluation broken? the incoherence of coherence.
\newblock \emph{Advances in neural information processing systems}, 34: 2018--2033.

\bibitem[{Ippolito et~al.(2022)Ippolito, Yuan, Coenen, and Burnam}]{ippolitoCreativeWritingAIPowered2022}
Ippolito, D.; Yuan, A.; Coenen, A.; and Burnam, S. 2022.
\newblock Creative {{Writing}} with an {{AI-Powered Writing Assistant}}: {{Perspectives}} from {{Professional Writers}}.
\newblock arXiv:2211.05030.

\bibitem[{Joseph and Morgan(2020)}]{joseph2020word}
Joseph, K.; and Morgan, J.~H. 2020.
\newblock When Do Word Embeddings Accurately Reflect Surveys on Our Beliefs about People?
\newblock In \emph{Proceedings of the 58th Annual Meeting of the Association for Computational Linguistics}, 4392--4415.

\bibitem[{Kaplan and Haenlein(2010)}]{kaplanUsersWorldUnite2010}
Kaplan, A.~M.; and Haenlein, M. 2010.
\newblock Users of the World, Unite! {{The}} Challenges and Opportunities of {{Social Media}}.
\newblock \emph{Business Horizons}, 53(1): 59--68.

\bibitem[{Katz and Rice(2002)}]{katz2002social}
Katz, J.~E.; and Rice, R.~E. 2002.
\newblock \emph{Social consequences of Internet use: Access, involvement, and interaction}.
\newblock MIT press.

\bibitem[{Kim et~al.(2024)Kim, Han, Adar, Kay, and Chung}]{kimAuthorsValuesAttitudes2024}
Kim, T.; Han, H.; Adar, E.; Kay, M.; and Chung, J. J.~Y. 2024.
\newblock Authors' {{Values}} and {{Attitudes Towards AI-bridged Scalable Personalization}} of {{Creative Language Arts}}.
\newblock In \emph{Proceedings of the 2024 {{CHI Conference}} on {{Human Factors}} in {{Computing Systems}}}, {{CHI}} '24, 1--16. New York, NY, USA: Association for Computing Machinery.
\newblock ISBN 9798400703300.

\bibitem[{Laestadius et~al.(2024)Laestadius, Bishop, Gonzalez, Illen{\v c}{\'i}k, and {Campos-Castillo}}]{laestadiusTooHumanNot2024}
Laestadius, L.; Bishop, A.; Gonzalez, M.; Illen{\v c}{\'i}k, D.; and {Campos-Castillo}, C. 2024.
\newblock Too Human and Not Human Enough: {{A}} Grounded Theory Analysis of Mental Health Harms from Emotional Dependence on the Social Chatbot {{Replika}}.
\newblock \emph{New Media \& Society}, 26(10): 5923--5941.

\bibitem[{Lamerichs(2023)}]{lamerichsGenerativeAINext2023}
Lamerichs, N. 2023.
\newblock Generative {{AI}} and the {{Next Stage}} of {{Fan Art}}.

\bibitem[{Lamerichs and Ossa(2023)}]{lamerichsFandomAlgorithmPrompting2023}
Lamerichs, N.; and Ossa, V. 2023.
\newblock Fandom, Algorithm, Prompting: {{Reconsidering}} Webcomics.
\newblock \emph{Studies in Comics}, 14(1): 137--149.

\bibitem[{Landis and Koch(1977)}]{landis1977measurement}
Landis, J.~R.; and Koch, G.~G. 1977.
\newblock The measurement of observer agreement for categorical data.
\newblock \emph{biometrics}, 159--174.

\bibitem[{Laufer(2025)}]{laufer2025ai}
Laufer, D. 2025.
\newblock \emph{AI love you. Gender and intimacy in user content regarding AI chatbot characters from Character. ai}.
\newblock Master's thesis, Univerzita Karlova.

\bibitem[{Ling(2004)}]{ling2004mobile}
Ling, R. 2004.
\newblock \emph{The mobile connection: The cell phone's impact on society}.
\newblock Elsevier.

\bibitem[{Liu, Pataranutaporn, and Maes(2024)}]{liuChatbotCompanionshipMixedMethods2024}
Liu, A.~R.; Pataranutaporn, P.; and Maes, P. 2024.
\newblock Chatbot {{Companionship}}: {{A Mixed-Methods Study}} of {{Companion Chatbot Usage Patterns}} and {{Their Relationship}} to {{Loneliness}} in {{Active Users}}.
\newblock arXiv:2410.21596.

\bibitem[{Maples et~al.(2024)Maples, Cerit, Vishwanath, and Pea}]{maplesLonelinessSuicideMitigation2024}
Maples, B.; Cerit, M.; Vishwanath, A.; and Pea, R. 2024.
\newblock Loneliness and Suicide Mitigation for Students Using {{GPT3-enabled}} Chatbots.
\newblock \emph{npj Mental Health Research}, 3(1): 1--6.

\bibitem[{McInnes et~al.(2017)McInnes, Healy, Astels et~al.}]{mcinnes2017hdbscan}
McInnes, L.; Healy, J.; Astels, S.; et~al. 2017.
\newblock HDBSCAN: Hierarchical Density Based Clustering.
\newblock \emph{J. Open Source Softw.}, 2(11): 205.

\bibitem[{McInnes, Healy, and Melville(2018)}]{mcinnes2018umap}
McInnes, L.; Healy, J.; and Melville, J. 2018.
\newblock UMAP: Uniform Manifold Approximation and Projection for Dimension Reduction.
\newblock \emph{arXiv preprint arXiv:1802.03426}.

\bibitem[{Milli and Bamman(2016)}]{milliCanonicalTextsComputational2016}
Milli, S.; and Bamman, D. 2016.
\newblock Beyond Canonical Texts: {{A}} Computational Analysis of Fanfiction.
\newblock In \emph{Proceedings of the 2016 {{Conference}} on {{Empirical Methods}} in {{Natural Language Processing}}}, 2048--2053.

\bibitem[{Monroe, Colaresi, and Quinn(2008)}]{monroe2008fightin}
Monroe, B.~L.; Colaresi, M.~P.; and Quinn, K.~M. 2008.
\newblock Fightin' Words: Lexical Feature Selection and Evaluation for Identifying the Content of Political Conflict.
\newblock \emph{Political Analysis}, 16(4): 372--403.

\bibitem[{Munger(2019)}]{munger2019limited}
Munger, K. 2019.
\newblock The Limited Value of Non-Replicable Field Experiments in Contexts With Low Temporal Validity.
\newblock \emph{Social Media+ Society}, 5(3): 2056305119859294.

\bibitem[{Nguyen et~al.(2024)Nguyen, Taher, Hong, Possobom, Gopalakrishnan, Raj, Li, Soled, Birnbaum, Kumar, and Choudhury}]{nguyenLargeLanguageModels2024}
Nguyen, V.~C.; Taher, M.; Hong, D.; Possobom, V.~K.; Gopalakrishnan, V.~T.; Raj, E.; Li, Z.; Soled, H.~J.; Birnbaum, M.~L.; Kumar, S.; and Choudhury, M.~D. 2024.
\newblock Do {{Large Language Models Align}} with {{Core Mental Health Counseling Competencies}}?
\newblock arXiv:2410.22446.

\bibitem[{Ouyang et~al.(2023)Ouyang, Wang, Liu, Zhong, Jiao, Iter, Pryzant, Zhu, Ji, and Han}]{ouyangShiftedOverlookedTaskoriented2023}
Ouyang, S.; Wang, S.; Liu, Y.; Zhong, M.; Jiao, Y.; Iter, D.; Pryzant, R.; Zhu, C.; Ji, H.; and Han, J. 2023.
\newblock The {{Shifted}} and {{The Overlooked}}: {{A Task-oriented Investigation}} of {{User-GPT Interactions}}.
\newblock arXiv:2310.12418.

\bibitem[{Pentina, Hancock, and Xie(2023)}]{pentinaExploringRelationshipDevelopment2023}
Pentina, I.; Hancock, T.; and Xie, T. 2023.
\newblock Exploring Relationship Development with Social Chatbots: {{A}} Mixed-Method Study of Replika.
\newblock \emph{Computers in Human Behavior}, 140: 107600.

\bibitem[{Popova(2018)}]{popovaArrangedMarriage2018}
Popova, M. 2018.
\newblock {{Rewriting}} the {{Romance}}: {{Emotion Work}} and {{Consent}} in {{Arranged Marriage Fanfiction}}.
\newblock \emph{Journal of Popular Romance Studies}, 7.

\bibitem[{Roose(2024)}]{rooseCanAIBe2024}
Roose, K. 2024.
\newblock Can {{A}}.{{I}}. {{Be Blamed}} for a {{Teen}}'s {{Suicide}}?
\newblock \emph{The New York Times}.

\bibitem[{Schnoebelen et~al.(2022)Schnoebelen, Silge, Hayes, and Silge}]{schnoebelen2022package}
Schnoebelen, T.; Silge, J.; Hayes, A.; and Silge, M.~J. 2022.
\newblock Package ‘tidylo’.

\bibitem[{Skjuve et~al.(2021)Skjuve, F{\o}lstad, Fostervold, and Brandtzaeg}]{skjuveMyChatbotCompanion2021}
Skjuve, M.; F{\o}lstad, A.; Fostervold, K.~I.; and Brandtzaeg, P.~B. 2021.
\newblock My {{Chatbot Companion}} - a {{Study}} of {{Human-Chatbot Relationships}}.
\newblock \emph{International Journal of Human-Computer Studies}, 149: 102601.

\bibitem[{Smith et~al.(2025)Smith, Shugars, Khanam, Mbonu, Lella, and Myers}]{smith2025black}
Smith, M.~A.; Shugars, S.; Khanam, S.; Mbonu, A.; Lella, O. S. K.~M.; and Myers, C.~L. 2025.
\newblock The Black Pill:(Re) conceptualizing the Black Right in the Era of YouTube Influencers.
\newblock \emph{Social Media+ Society}, 11(1): 20563051251329078.

\bibitem[{Song et~al.(2024)Song, Pendse, Kumar, and Choudhury}]{songTypingCureExperiences2024}
Song, I.; Pendse, S.~R.; Kumar, N.; and Choudhury, M.~D. 2024.
\newblock The {{Typing Cure}}: {{Experiences}} with {{Large Language Model Chatbots}} for {{Mental Health Support}}.
\newblock arXiv:2401.14362.

\bibitem[{Stieglitz et~al.(2018)Stieglitz, Brachten, Ross, and Jung}]{stieglitzSocialBotsDream2018}
Stieglitz, S.; Brachten, F.; Ross, B.; and Jung, A. 2018.
\newblock Do {{Social Bots Dream}} of {{Electric Sheep}}? {{A Categorisation}} of {{Social Media Bot Accounts}}.
\newblock In \emph{{{ACIS}} 2017 {{Proceedings}}}, 1--11.

\bibitem[{Traag, Waltman, and Van~Eck(2019)}]{traag2019louvain}
Traag, V.~A.; Waltman, L.; and Van~Eck, N.~J. 2019.
\newblock From Louvain to Leiden: guaranteeing well-connected communities.
\newblock \emph{Scientific reports}, 9(1): 1--12.

\bibitem[{Tu et~al.(2024)Tu, Fan, Tian, and Yan}]{tuCharacterEvalChineseBenchmark2024}
Tu, Q.; Fan, S.; Tian, Z.; and Yan, R. 2024.
\newblock {{CharacterEval}}: {{A Chinese Benchmark}} for {{Role-Playing Conversational Agent Evaluation}}.
\newblock arXiv:2401.01275.

\bibitem[{{Upton-Clark}(2024)}]{upton-clarkCharacterAIBeingSued2024}
{Upton-Clark}, E. 2024.
\newblock Character.{{AI}} Is Being Sued for Encouraging Kids to Self-Harm.

\bibitem[{Vasiliev(2020)}]{vasiliev2020natural}
Vasiliev, Y. 2020.
\newblock \emph{Natural language processing with Python and spaCy: A practical introduction}.
\newblock No Starch Press.

\bibitem[{Wang(2024)}]{wang2024eliza}
Wang, K. 2024.
\newblock From ELIZA to ChatGPT: A brief history of chatbots and their evolution.
\newblock \emph{Applied and Computational Engineering}, 39: 57--62.

\bibitem[{Xie, Pentina, and Hancock(2023)}]{xieFriendMentorLover2023}
Xie, T.; Pentina, I.; and Hancock, T. 2023.
\newblock Friend, Mentor, Lover: Does Chatbot Engagement Lead to Psychological Dependence?
\newblock \emph{Journal of Service Management}, 34(4): 806--828.

\bibitem[{Yang, Wu, and Hearst(2024)}]{yangHumanAIInteractionAge2024}
Yang, D.; Wu, S.~T.; and Hearst, M.~A. 2024.
\newblock Human-{{AI Interaction}} in the {{Age}} of {{LLMs}}.
\newblock In \emph{NAACL'24}, 34--38.

\bibitem[{Zhang et~al.(2025)Zhang, Zhao, Hancock, Kraut, and Yang}]{zhang2025rise}
Zhang, Y.; Zhao, D.; Hancock, J.~T.; Kraut, R.; and Yang, D. 2025.
\newblock The rise of AI companions: how human-chatbot relationships influence well-being.
\newblock \emph{arXiv preprint arXiv:2506.12605}.

\bibitem[{Zhao et~al.(2024)Zhao, Ren, Hessel, Cardie, Choi, and Deng}]{zhaoWildChat1MChatGPT2024}
Zhao, W.; Ren, X.; Hessel, J.; Cardie, C.; Choi, Y.; and Deng, Y. 2024.
\newblock {{WildChat}}: {{1M ChatGPT Interaction Logs}} in the {{Wild}}.
\newblock arXiv:2405.01470.

\bibitem[{Zheng et~al.(2023)Zheng, Chiang, Sheng, Li, Zhuang, Wu, Zhuang, Li, Lin, Xing, Gonzalez, Stoica, and Zhang}]{zhengLMSYSChat1MLargeScaleRealWorld2023}
Zheng, L.; Chiang, W.-L.; Sheng, Y.; Li, T.; Zhuang, S.; Wu, Z.; Zhuang, Y.; Li, Z.; Lin, Z.; Xing, E.~P.; Gonzalez, J.~E.; Stoica, I.; and Zhang, H. 2023.
\newblock {{LMSYS-Chat-1M}}: {{A Large-Scale Real-World LLM Conversation Dataset}}.
\newblock arXiv:2309.11998.

\end{thebibliography}
\end{document}